\documentclass[twocolumn,english,aps,preprint,prl,reprint,superscriptaddress,longbibliography]{revtex4-1}
\usepackage[all]{nowidow}
\usepackage[T1]{fontenc}
\usepackage[utf8]{inputenc}
\usepackage{amssymb}
\usepackage{graphicx}
\usepackage{natbib}
\usepackage{etoolbox}
\usepackage{scrextend}
\usepackage{placeins}
\makeatletter

\usepackage[usenames,dvipsnames]{color}
\usepackage{babel}

\DeclareMathAlphabet{\mathpzc}{OT1}{pzc}{m}{it}
\usepackage{comment}
\usepackage{verbatim}
\usepackage{amsmath}
\usepackage{mathtools}
\usepackage{xfrac}
\usepackage{bm}
\usepackage{xcolor}
\usepackage{color}
\definecolor{red}{rgb}{1,0,0}
\definecolor{green}{rgb}{0,1,0}
\definecolor{blue}{rgb}{0,0,1}
\definecolor{c1}{rgb}{0.178,0.63,0.17}
\definecolor{c2}{rgb}{0.09,0.740,0.81}
\definecolor{c3}{rgb}{0.12,0.46,0.7}
\definecolor{c4}{rgb}{0.58,0.40,0.74}
\let\saved@includegraphics\includegraphics
\AtBeginDocument{\let\includegraphics\saved@includegraphics}
\renewenvironment*{figure}{\@float{figure}}{\end@float}

\newcommand{\ket}[1]{\left|#1\right>}

\newcommand{\nqninet}{N_{q_\text{9}} \left( t \right)}

\newcommand{\rd}[1]{{\color{BrickRed}}}

\usepackage{pdfpages}
\makeatletter
\AtBeginDocument{\let\LS@rot\@undefined}
\makeatother

\begin{document}

\vspace{-0.5em}
{\onecolumngrid 
\centering{{\noindent \large \textbf{Direct measurement of non-local interactions in the many-body localized phase}}}

\vspace{0.5em}

\centering{
{\footnotesize
B. Chiaro$^{* 1}$,
C. Neill$^{* 2}$,
A. Bohrdt$^{* 3,4}$,
M. Filippone$^{* 5}$,
F. Arute$^2$,
K. Arya$^2$,
R. Babbush$^2$,
D. Bacon$^2$,
J. Bardin$^2$,
R. Barends$^2$,
S. Boixo$^2$,
D. Buell$^2$,
B. Burkett$^2$,
Y. Chen$^2$,
Z. Chen$^2$,
R. Collins$^2$,
A. Dunsworth$^2$,
E. Farhi$^2$,
A. Fowler$^2$,
B. Foxen$^2$,
C. Gidney$^2$,
M. Giustina$^2$,
M. Harrigan$^2$,
T. Huang$^2$,
S. Isakov$^2$,
E. Jeffrey$^2$,
Z. Jiang$^2$,
D. Kafri$^2$,
K. Kechedzhi$^2$,
J. Kelly$^2$,
P. Klimov$^2$,
A. Korotkov$^2$,
F. Kostritsa$^2$,
D. Landhuis$^2$,
E. Lucero$^2$,
J. McClean$^2$,
X. Mi$^2$,
A. Megrant$^2$,
M. Mohseni$^2$,
J. Mutus$^2$,
M. McEwen$^2$,
O. Naaman$^2$,
M. Neeley$^2$,
M. Niu$^2$,
A. Petukhov$^2$,
C. Quintana$^2$,
N. Rubin$^2$
D. Sank$^2$,
K. Satzinger$^2$,
A. Vainsencher$^2$,
T. White$^2$,
Z. Yao$^2$,
P. Yeh$^2$,
A. Zalcman$^2$,
V. Smelyanskiy$^2$,
H. Neven$^2$,
S. Gopalakrishnan$^6$,
D. Abanin$^7$,
M. Knap$^{3,4}$,
J. Martinis$^{1,2}$,
and 
P. Roushan$^2$}

\vspace{1em}

\begin{addmargin}[-0.1em]{-0.10em}
{\footnotesize
$^1${\textit{Department of Physics, University of California, Santa Barbara, CA, USA}}
$^2${\textit{Google Inc., Santa Barbara, CA, USA}}
$^3${\textit{Department of Physics and Institute for Advanced Study, Technical University of Munich, Germany}}
$^4${\textit{Munich Center for Quantum Science and Technology, M{\"u}nchen, Germany}}
$^5${\textit{Department of Quantum Matter Physics, University of Geneva, Switzerland }}
$^6${\textit{Department of Physics and Astronomy, College of Staten Island, USA}}
$^7${\textit{Department of Theoretical Physics, University of Geneva, Switzerland}}
}
\end{addmargin}
}}
\vspace{-0.5em}

\setcounter{topnumber}{2} 
\setcounter{bottomnumber}{2} 
\setcounter{totalnumber}{4} 
\renewcommand{\topfraction}{0.99} 
\renewcommand{\bottomfraction}{0.99} 
\renewcommand{\textfraction}{0} 
\renewcommand{\floatpagefraction}{0.999} 
\setlength{\floatsep}{5pt plus 2pt minus 2pt}  
\setlength{\textfloatsep}{5pt plus 2pt minus 2pt} 
\setlength{\intextsep}{5pt plus 2pt minus 2pt}  

\begin{abstract}
\begin{addmargin}[-5.8em]{-0.4em}
\textbf{The interplay of interactions and strong disorder can lead to an exotic quantum many-body localized (MBL) phase. Beyond the absence of transport, the MBL phase has distinctive signatures, such as slow dephasing and logarithmic entanglement growth; they commonly result in slow and subtle modification of the dynamics, making their measurement challenging. Here, we experimentally characterize these properties of the MBL phase in a system of coupled superconducting qubits. By implementing phase sensitive techniques, we map out the structure of local integrals of motion in the MBL phase. Tomographic reconstruction of single and two qubit density matrices allowed us to determine the spatial and temporal entanglement growth between the localized sites. In addition, we study the preservation of entanglement in the MBL phase. The interferometric protocols implemented here measure affirmative correlations and allow us to exclude artifacts due to the imperfect isolation of the system. By measuring elusive MBL quantities, our work highlights the advantages of phase sensitive measurements in studying novel phases of matter. }
\end{addmargin}
\end{abstract}

\maketitle

\section{} 
\vspace{-10mm}

Disorder-induced localization is a ubiquitous phenomenon that occurs in both classical and quantum systems. In 1958 Anderson showed that in non-interacting systems disorder can change the structure of electronic wave-functions from being extended to exponentially localized\,\cite{Anderson1958}. This localized phase has been observed for systems of non-interacting phonons, photons, and matter-waves\,\cite{The50years,Billy2008, Weaver1990, Wiersma1997, Schwartz2007}.  The conventional wisdom had long been that systems of interacting particles do not localize and ultimately reach thermal equilibrium regardless of the disorder magnitude. However, recent work shows that localization may persist even in the presence of interactions between particles, establishing the many-body localized (MBL) phase as a robust, non-ergodic phase of quantum matter at finite temperature\,\cite{Basko2006, Gornyi2005, ImbriePRL2016, BlochMBL2015, demarco2015, Monroe2015}. 

The foremost characteristic of the MBL phase is the absence of transport and local relaxation to a thermal state~\cite{Basko2006, BlochMBL2015, demarco2015, GrossScience2016, Bordia2017, Leuschen2017, Roushan2018,Zha2020}; from this perspective, the MBL phase resembles a noninteracting Anderson insulator. But the dynamics of quantum information in the MBL phase are richer than in an Anderson insulator~\cite{Huse2007, Bardarson2012, Serbyn2013b, Huse2014, Serbyn2013, KnapPRL2014, Antonello2017, Serbyn2014, Yasaman2015,  Gopalakrishnan2015, Altman2015, ImbriePRL2016}. The two phases share the property of having extensively many spatially local integrals of motion. However, in the MBL phase, the integrals of motion interact in ways that lead to slow dephasing and the logarithmic growth of entanglement, among other consequences, some of which have been experimentally observed~\cite{Rispoli2019, Lukin2019, Brydges2019}. Directly probing the structure of these integrals of motion, which define the MBL phase, has proven experimentally challenging, as it is best accomplished with phase sensitive measurements.

\begin{figure}[h!] 
\centering
\includegraphics[width=80mm]{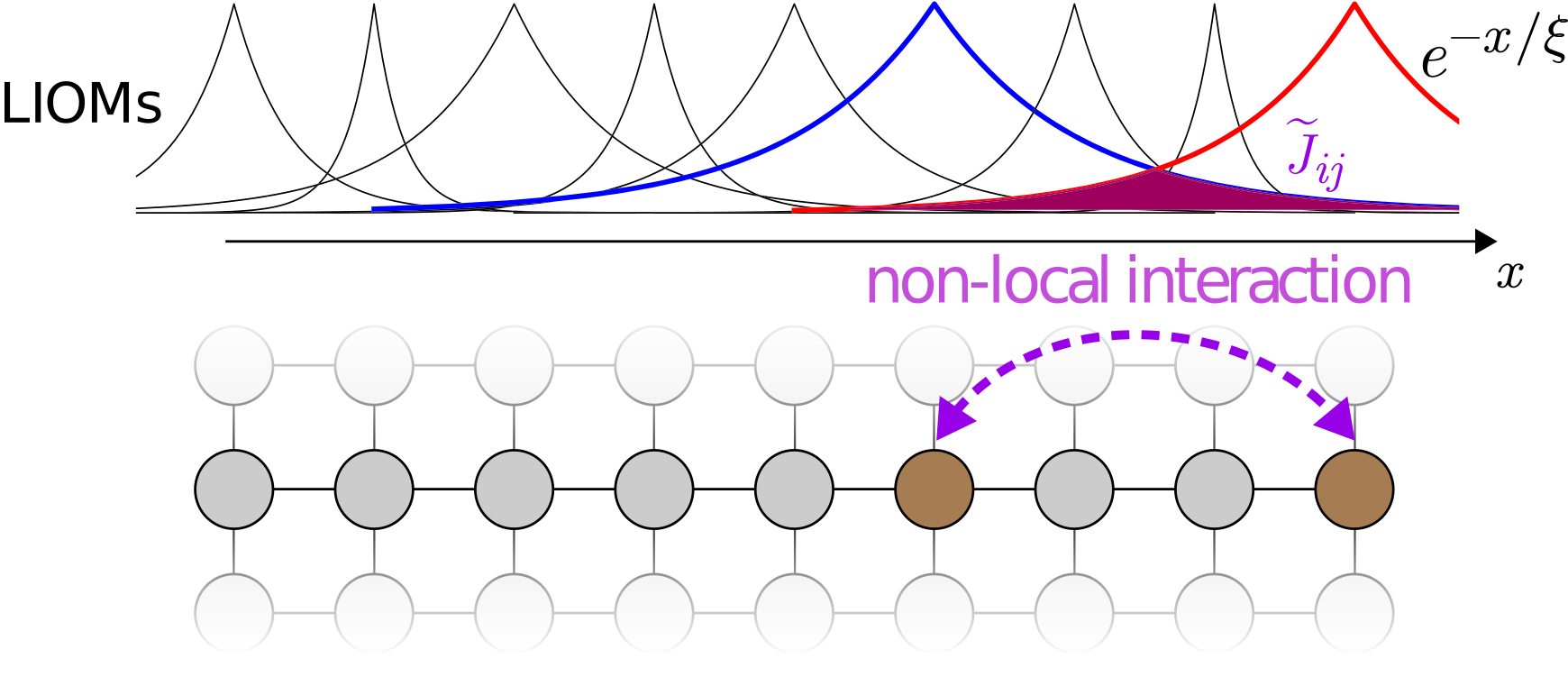}
\vspace{-0.9em}
\caption{ \textbf{Many-body localization with superconducting qubits.} The constituents of the many-body localized phase are localized orbitals (local integrals of motion, LIOMs). The LIOM structure is determined by the potential landscape, with greater disorder further localizaing the LIOMs, and hence decreasing their typical length scale $\xi$. The spatial disorder yields a distribution of the length scales $\xi$.  The shaded region indicates effective non-local interactions $\widetilde{J}_{ij}$ between two LIOMs, giving rise to non-trivial dephasing dynamics and logarithmic entanglement growth.
}
\end{figure}

\begin{figure*}[t] 
\centering
\includegraphics[width=178mm]{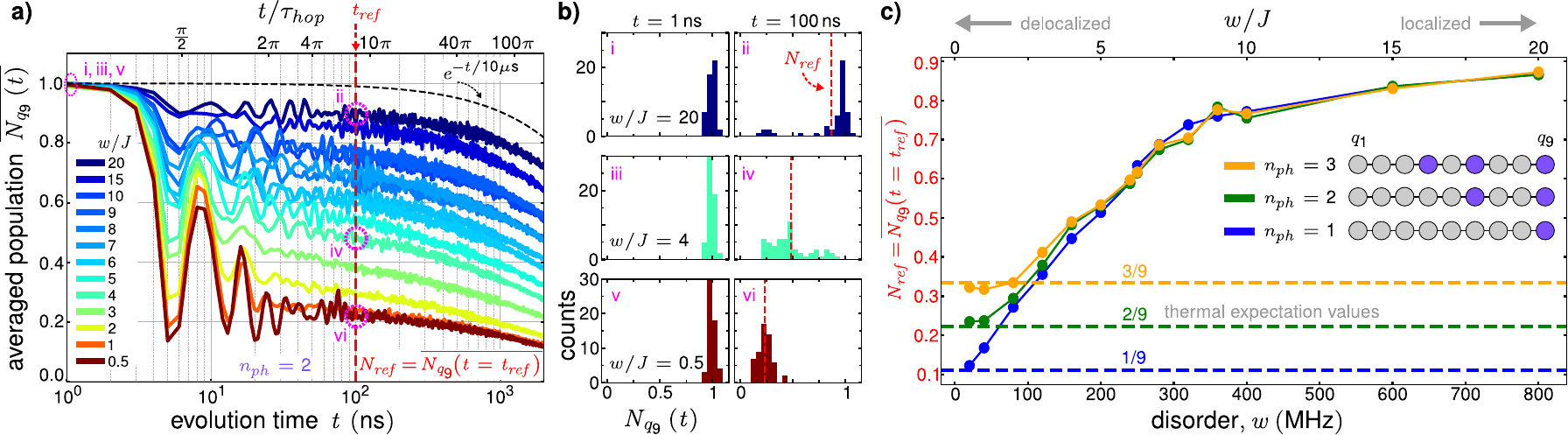}
\vspace{-0.7em}
\caption{\small
\textbf{Ergodicity breakdown at strong disorder.} \textbf{(a)} Disorder averaged on-site population vs. time for $n_{ph}=2$. In a chain of 9 qubits, two qubits were excited (q6, q9). The on-site population of q9 was measured for various magnitudes of disorder $w/J$, with $J=40\,\text{MHz}$ (averaged over 50 realizations). The parameter $\tau_{hop}=\left(2 \pi J \right)^{-1}$ has been introduced to connect the laboratory time $t$ with the hopping energy. $N_{ref}$ is defined to be the average on-site population across instances of disorder at the reference time $t_{ref}=100\,\text{ns}$, after initial transients have been damped. The dashed black line indicates average photon loss for a single qubit measured in isolation.  \textbf{(b)} Histograms of $N_{q9}(t)$
at the times and disorders indicated in (a) by numerals i - vi. \textbf{(c)} $N_{ref}$ vs. disorder for $n_{ph}=1, 2, 3$. Inset shows which qubits were initially excited. }
\vspace{-1em}
\end{figure*}

Using an array of coupled superconducting qubits in one- and two-dimensions, we study the dynamics of interacting photon excitations in a disordered potential. The Hamiltonian is described by the Bose-Hubbard model 
\begin{eqnarray}
H_\text{BH} &=& \underbrace{\sum\limits_{i}^{n_Q} h_i a^{\dagger}_{i}a_{i}}_{\text{on-site detuning}} +
\underbrace{\frac{U}{2}\sum\limits_{i}^{n_Q} \,a^{\dagger}_{i}a_i(a^{\dagger}_{i}a_i-1)}_{\text{Hubbard interaction}} \nonumber \\
&+& \underbrace{J \sum_{\left< i, j \right>}\left( a^{\dagger}_{i}a_j+\text{h.c.}\right) }_{\text{NN coupling / hopping}},
\end{eqnarray}
\noindent
where $a^{\dagger}$ ($a$) denotes the bosonic creation (annihilation) operator, $h_i\in \left[ -w, w \right]$ is the random on-site detuning drawn from a uniform distribution of width $2w$, $J$ is the hopping rate between nearest neighbor lattice sites, $U$ is the on-site Hubbard interaction, and $n_Q$ is the number of qubits; see supplementary material for details on the characterization\,\cite{supplement}. The qubit frequency, the nearest neighbor coupling, and the nonlinearity set $h_i$, $J$, and $U$, respectively. We are able to tune $h_i$ and $J$ independently at a fixed nonlinearity $U=160$ MHz.

The localized regime of Eqn.\,(1) is obtained when the frequency detunings $h_i$ are large compared to $J$. In this regime, the eigenstates of the Hamiltonian are product states of localized orbitals, referred to as local integrals of motion (LIOMs), which are nearly qubit states but have a spatial extent that decays exponentially across the neighboring qubits\,(Fig.\,1). Before measuring the properties of the LIOMs, we show that our system of qubits is manifestly localized by studying the conventional relaxation dynamics.

\section{} 
\vspace{-16mm}

\textcolor{blue}{ Evidence for the breakdown of ergodic dynamics} can be obtained by measuring the mobility of excitations in a 1$\times$9 qubit array. In Fig.\,2 we initialize the system with a number of photon excitations $n_{ph}$ by preparing $1$, $2$, or $3$ qubits in the single excitation Fock state. We measure the population on one of the initially excited qubits as the system evolves under Hamiltonian\,(1). The disorder averaged population at $q_9$ (the observation site) $\overline{ \nqninet }$ for $n_{ph}=2$ is shown in panel (a). We choose a reference time $t_{ref}$, in which $\overline{ \nqninet }$ approaches an asymptotic value after initial transients have been damped, but before the dynamics of our system are dominated by relaxation or dephasing at large time scales (dashed black line)\,\cite{supplement, Znidaric2015, Levi2016, Fischer2016, Luschen2017, vanNieuwenburg2017}. 

The distribution of $\nqninet$ for selected disorder magnitudes at $t=1\,\text{ns}$ and $t=t_{ref}$ are shown in Fig.\,2(b). At $t=1\,\text{ns}$ the excitations have not propagated, and there is a tight distribution close to the initial values, regardless of the value of disorder. At $t=t_{ref}$ the distribution is narrow for low disorder and becomes wider with tails at larger disorders. This can be understood because at high disorder, level resonances are increasingly rare which inhibits mobility. The tail of the distribution results from these rare cases. At low disorder, excitations can propagate freely between qubits and the behavior of each disorder instance is typical, giving rise to narrow distributions.

\begin{figure}[t]
\includegraphics[width=80mm]{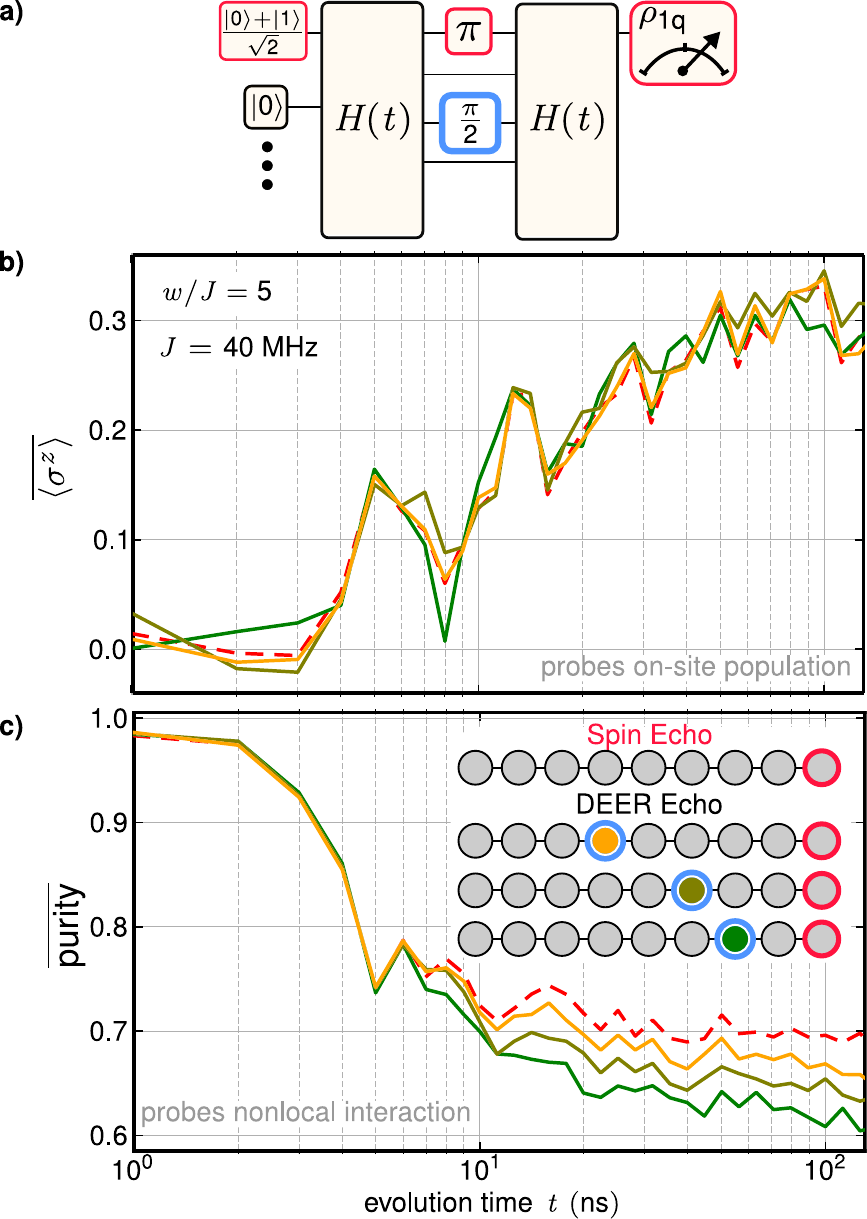}
\vspace{-0.7em}
\caption{\small
\textbf{Interferometric signatures of remote entanglement.}
\textbf{(a)} SE and DEER pulse sequences. DEER differs from SE by the addition of a remote $\pi\, /\,2$-pulse simultaneous with the SE $\pi$-pulse between the free precession intervals. \textbf{(b)}  $\overline{ \left< \sigma^z \right> } =\overline{ \left< 1 - 2 a^\dagger a \right> }$, and \textbf{(c)}  purity of the single qubit for SE (red dashed) and DEER (solid) experiments. The remote DEER pulse induces dephasing, decreasing the purity. The contrast between SE and DEER probes the non-local interaction $\widetilde{J}_{ij}$ between the SE lattice site and the DEER site.}
\end{figure}

Fig.\,2(c) shows the disorder averaged population at $t_{ref}=100\, \text{ns}$ as a function of the disorder strength. 
At weak disorder our observations are consistent with the ergodic hypothesis that each of the accessible photon states is equally likely to be observed. A uniform averaging over the available phase space implies that the  occupancy of a given qubit should be $n_{ph}\,/n_Q$. 
However, as we increase the disorder strength, significant deviations from the thermal value are observed, which indicates that system becomes many-body localized. We note that with more photons in the system, the population converges to its thermal expectation value at higher disorders. This is expected because the increased interactions assist with the thermalization process and drive delocalization. 
In the case of a single excitation our system is non-interacting and hence localized for all disorder magnitudes. The apparent approach of the population to the thermal value at extremely weak disorder indicates the regime where the single-particle localization length exceeds our system size. In two spatial dimensions, we observe similar signatures for localization, see supplement~\cite{supplement}. 

\section{} 
\vspace{-20mm}

\textcolor{blue}{Nonlocal interactions between the LIOMs are a defining characteristic of the MBL state.}  As the system is localized, the nontrivial dynamics of the system are expressed in terms of phase relationships between the LIOMs, which are most naturally observed observed through phase sensitive protocols rather than measurements of population.  In the localized regime, Eqn.\,(1) can be brought into a diagonal form by a finite set of local unitary transformations\,\cite{Serbyn2013, Huse2014}. In this basis there is no hopping and the Hamiltonian can be written in terms of on-site detunings and non-local interactions,
\vspace{10pt}
\begin{equation} 
\tilde{H}_{\tau} = \underbrace{\sum_i \widetilde{h}_i \tau^z_i}_{\text{on-site detuning}} + \underbrace{ \sum_{i,j} \widetilde{J}_{ij}\tau^z_i\tau^z_j + \sum_{ijk} \widetilde{J}_{ijk}\tau^z_i\tau^z_j\tau^z_k + \mathellipsis.}_{\text{non-local interaction}}
\end{equation}
\noindent    
The $\tau^z_j$ are Pauli operators that commute with $\tilde H_{\tau}$ and are hence conserved; the system is localized. Note that in the parameter regime considered here, two and more excitations per qubit occur only virtually\,\cite{supplement}.
The non-local interactions $\widetilde{J}_{ij}$, $\widetilde{J}_{ijk}$, $\ldots$ generate entanglement throughout the localized system and can be unambiguously established by adopting interferometric methods inspired by NMR protocols\,\cite{KnapPRL2014}, which are closely related to measuring out-of-time-order correlators. 

Fig.\,3(a) illustrates a conventional spin-echo (SE) sequence and its extension double electron-electron resonance echo (DEER) which we use to provide a differential measurement of phase accumulation with and without a remote perturbation. The construction and effects of these pulse sequences can be understood from Eqn.\,(2). Deep in the MBL phase, the LIOMs are nearly localized on individual qubits. The SE $\pi$-pulse between free precession intervals essentially negates the local frequency detuning, reversing the evolution and hence reversing phase accumulation. The role of the additional $\pi/2$-pulse in the DEER sequence is to make the SE refocusing incomplete by an amount depending on the nonlocal interaction. Thus the technique directly probes the strength of this non-local interaction\,\cite{supplement}.  A comparison with closed system numerics is presented in the supplement. 

The measurement of on-site population, depicted in panel (b), shows that the remote $\pi / 2$-pulse in the DEER sequence does not alter the population on the observation site, assuring that the system is in the localized regime.

Comparing the difference of SE and DEER\,(panel (c)), we see that the additional \emph{differential} relaxation in the DEER case is a pure interference effect that directly measures the non-local interaction between distant localized sites. In addition, the difference between SE and DEER decreases as the distance between the SE site and remote disturbance site is increased. This can be understood from the decaying nature of the interactions between the LIOMs with distance. The interferometric protocol is thus demonstrating the foundational interaction effects of MBL states. As a next step, we characterize the distribution of the couplings $\widetilde{J}_{ij}$. 

\begin{figure}[t]
\includegraphics[width=80mm]{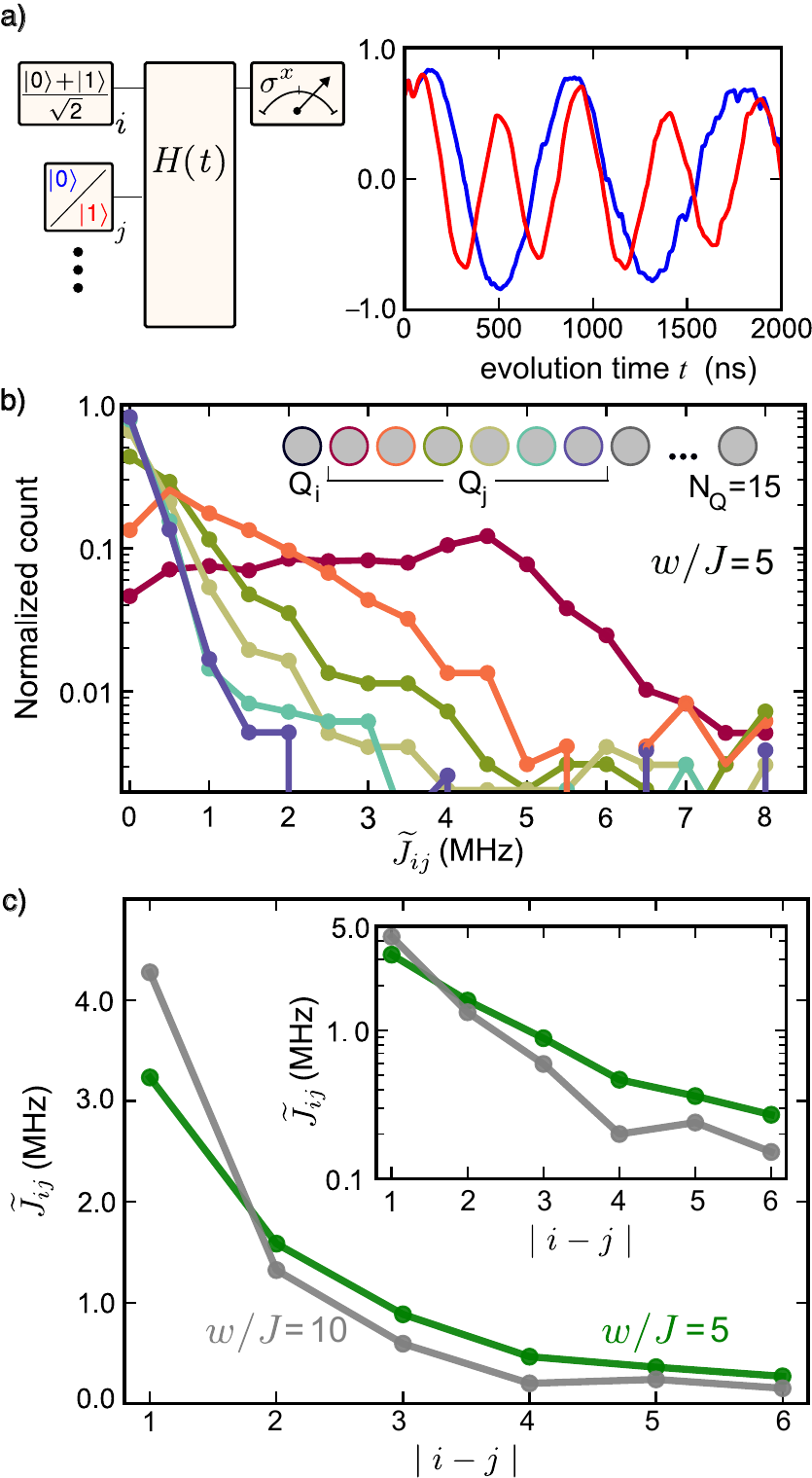}
\vspace{-0.7em}
\caption{\small
\textbf{Distribution of the couplings between the localized orbitals} 
\textbf{(a)} The pulse sequence for measuring $\widetilde{J}_{ij}$ showing that the evolution of $\langle \sigma^X_i \rangle$ on $Q_i$ for $Q_j$ initialized in $\ket{0}$ (red) and in $\ket{1}$ (blue). \textbf{(b)} The histogram of $\widetilde{J}_{ij}$ values measured for 1000 instances of disorder vs. distance between $Q_i$ and $Q_j$ . \textbf{(c)} The disorder-instance-averaged values of $\widetilde{J}_{ij}$ in linear (main) and semi-logarithmic scale(inset) for two ratios of disorder $w/J=5$ (green) and  $w/J=5$ (gray).}    
\end{figure}

\begin{figure*}[t] 
\includegraphics[width=160mm]{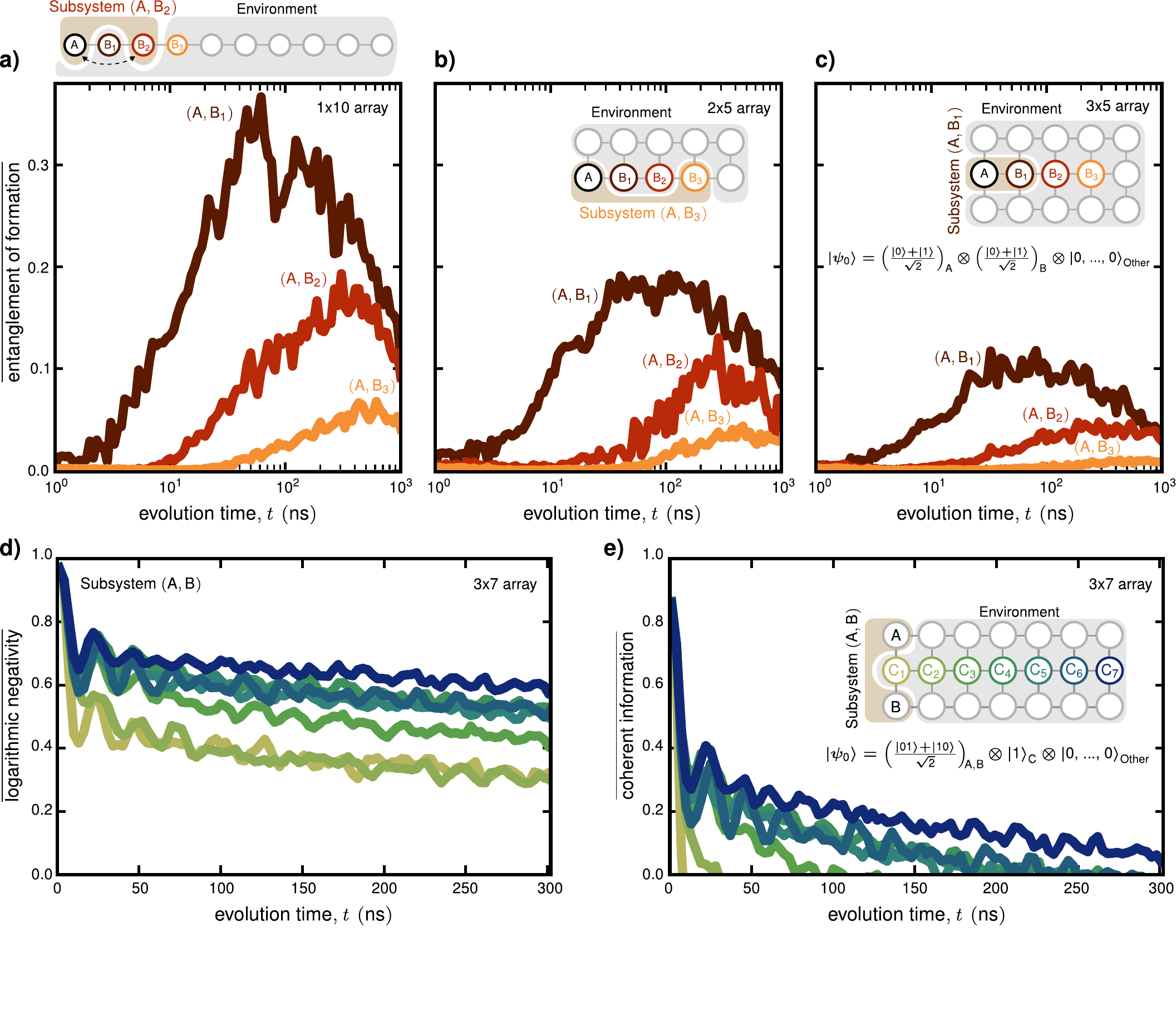}
\vspace{-4.5em}

\caption{\small \textbf{Growth and preservation of entanglement between localized sites.}
Entanglement of formation between qubits in various 2-qubit subsystems (A,B$_i$). To observe the development of entanglement between sites A and B the subsystem is initialized in a product of single qubit superposition states and the entanglement of formation of the two qubit density matrix is extracted, for subsystems of \textbf{(a)} $1\times10$, \textbf{(b)} $2\times5$, and \textbf{(c)} $3\times5$ array of qubits with $J=30$\,MHz and $w/J=10$. In a 2 qubit subsystem (A,B) of a 3 by 7 array of qubits, a Bell pair is created, and the Logarithmic negativity \textbf{(d)} and coherent information \textbf{(e)} are extracted from measurements of the subsystem density matrix and averaged over 80 realizations of disorder for $J=30$\,MHz with $w/J=12$.  We initialize the environment with an excitation at a position $C_i$ which is varied.}
\end{figure*}


\textcolor{blue}{A hallmark of the MBL phase} are the non-local interactions $\widetilde{J}_{ij}$ between LIOMs. To investigate the emergent non-local interactions, we measure the distribution of the couplings $P(\widetilde{J}_{ij})$\,(Fig.\,4) with a conditional phase measurement made possible by our ability to drive on site rotations. As shown in panel\,(a), our protocol consists of preparing qubit $Q_i$ in a superposition state $(\ket{0} + \ket{1})/\sqrt2$ and then measure evolution of its $\langle \sigma^X_i \rangle$ under two conditions: when $Q_j$ is in $\ket{0}$ state and when it is in $\ket{1}$ state. The rate of phase accumulation of $Q_i$ is conditioned on the state of $Q_j$ and thereby permits the extraction of the $\widetilde{J}_{ij}$.  Experimentally, we measure the dominant low frequency peak and associate its shift as the $\widetilde{J}_{ij}$. Repeating this process several times for different disorder realizations, we obtain the distribution of the couplings $\widetilde{J}_{ij}$. We find the $\widetilde{J}_{ij}$ to be broadly distributed, Fig. 4\,(b), with a mean that is rapidly decaying with increasing distance between the qubits (panel\,(c)). The broad distribution of the couplings has profound consequences. In particular, upon disorder averaging the entanglement entropy between the two qubits grows logarithmically in time and saturates at a finite value. By contrast a sharp distribution of the couplings would lead to an oscillatory behavior of the entanglement entropy between two entities\,\cite{Serbyn2013b}. 

\section{} 
\vspace{-15mm}
\textcolor{blue}{We investigate the formation and preservation of entanglement} between two qubits A and B that are embedded in an MBL environment as illustrated in Fig.\,5(a). Details of the 2D device used for these measurements can be found in \cite{Arute2019}. The entanglement of formation (EOF) is a proxy for the entanglement cost, i.e., the amount of entanglement directly between qubits A and B that would be required to asymptotically produce the observed two-qubit mixed state density matrix~\cite{Wootters1998}. We emphasize that because we are affirmatively detecting a quantum correlation between sites of the subsystem, the observed EOF cannot be attributed to open system effects which would tend to suppress the correlations.
The EOF is therefore a more conservative entanglement measure than conventional measures, such as the von Neumann entanglement entropy, and a valuable tool for characterizing realistic experimental systems, which are always coupled to environmental degrees of freedom.

In panels (a) to (c) we initialize the subsystem in a product state of single qubit superpositions and observe the development of entanglement between the subsystem qubits. Regardless of geometry of the qubit array, entanglement grows gradually between the localized, spatially separated sites over several hopping times. The entanglement grows faster when the subsystem qubits are closer to each other. This can be understood by considering two isolated qubits, which are becoming correlated with a rate given by the effective interactions $\widetilde{J}_{ij}$ that increases with decreasing distance (Fig.\,4(c)). The EOF for a single disorder realization possesses a sinusoidal shape. However, due to the disorder average over the broad distribution of the couplings $\widetilde{J}_{ij}$ the EOF saturates at intermediate times and only decays at late times due to open system effects. The EOF results have to be contrasted with the von Neumann entanglement entropy, which would continuously increase because it includes entanglement with all degrees of freedom external to the subsystem.

As the system geometry is transformed from 1D to 2D (panels (a) to (c)) there is an overall trend of suppressed EOF.  This is because of the monogamy of entanglement principle\,\cite{Wootters2000}. Compared with 1D, in 2D each qubit has additional neighbors, which changes the structure of the LIOMs and provides more transport channels, thus enhancing the spread of entanglement.  The monogamic principle states that there is a maximum degree to which two qubits may be correlated, and that entangling (correlating) either member of this pair with other qubits necessarily decorrelates the first two.  Thus in the higher dimensional systems shown here the subsystem qubits entangle with the environmental qubits to a greater extent thereby reducing the degree to which the subsystem qubits can be correlated.

At long times, the interaction between subsystem qubits is outcompeted by the interaction of the subsystem with the environmental qubits and the open system and the EOF declines.  We highlight that in our system the EOF, an affirmative correlation measure, detects correlations between sites with a large separation, e.g. (A, B$_3$) even though they are embedded in a large system of qubits which are never truly isolated.

The results thus far illustrate how interaction effects propagate entanglement throughout the system.  However, because MBL systems are non-thermal, features of their initial state remain imprinted on them.  Stable non-thermal local occupations exemplify this behavior, however the extension of this memory to quantum correlations has not been demonstrated experimentally.
To probe this aspect, we prepare a maximally entangled Bell state between two subsystem qubits in a 3$\times$7 qubit array and monitor the subsystem density matrix as the pair interacts with a remote photon. We focus on the distillable entanglement (DE), i.e., the entanglement which can be extracted from the mixed density matrix.  The upper and lower bounds of the DE are the logarithmic negativity entropy and the coherent information entropy respectively, shown in Fig.\,5(d) and (e).

The initial drop of DE, on the single hopping timescale, is attributed to population transfer from the Bell pair into the environmental qubits. Thereafter, interaction with the remote photon induces local dephasing in the subsystem, decorrelating the subsystem qubits according to the monogamy of entanglement principle.  With the remote photon at larger distances, the DE remains finite over several hopping times, in contrast to the behavior at low disorder, c.f. supplemental material.  The entanglement is increasingly disturbed as the remote photon is brought closer to the Bell pair and the coherent information that lower bounds the DE approaches zero at earlier times.  This data illustrates that in the MBL phase, a memory of the initial quantum correlations persists to late times and indicates that the excitation density is a critical parameter for this application.

By introducing phase sensitive algorithms and measurement, we have directly probed the nonlocal interactions responsible for entanglement propagation and mapped the spatial structure of the localized orbitals. The techniques introduced here extend easily to the characterization of digital algorithms and also more broadly to other synthetic quantum systems, thus offering a new toolkit to experimentally probe entanglement dynamics in a variety of settings.

\vspace{1em}
\noindent \textbf{Correspondence and requests for materials}
\small{should be addressed to P. Roushan\,(pedramr@google.com).}

\vspace{1em}
\noindent \textbf{Data availability}
Data that support the findings of this study are available from the corresponding authors upon reasonable request.

\vspace{1em}
\noindent * These authors contributed equally to this work.

\vspace{1em}
\noindent \textbf{Acknowledgments} \footnotesize{ The authors acknowledge valuable conversations with Jens Eisert and Andrew Daley. M.K. and A.B. acknowledge support from the Technical University of Munich - Institute for Advanced Study, funded by the German Excellence Initiative and the European Union FP7 under grant agreement 291763, the Deutsche Forschungsgemeinschaft (DFG, German Research Foundation) under Germany's Excellence Strategy--EXC-2111--390814868, DFG grant Nos. KN1254/1-1, KN1254/1-2 and DFG TRR80 (Project F8), the European Research Council (ERC) under the European Union's Horizon 2020 research and innovation programme (grant agreement No 851161).  M.F. also acknowledges support from the FNS/SNF Ambizione Grant PZ00P2\_174038. S.G. acknowledges support from NSF Grant No. DMR-1653271. M.K., A.B., D.A., and M. F. acknowledge support through Google Quantum NISQ award.}

\bibliography{MBL_Refs}
\bibliographystyle{naturemag}


\bigskip
\onecolumngrid
\newpage


\section*{Supplementary material}

# Supplemental information for "Direct measurement of non-local interactions in the many-body localized phase"

B. Chiaro*[1], C. Neill*[2], A. Bohrdt*[3,4], M. Filippone*[5], F. Arute[2], K. Arya[2], R. Babbush[2], D. Bacon[2], J. Bardin[2], R. Barends[2], S. Boixo[2], D. Buell[2], B. Burkett[2], J. Chen[2], Y. Chen[2], R. Collins[2], A. Dunsworth[2], E. Farhi[2], A. Fowler[2], B. Foxen[1], C. Gidney[2], M. Giustina[2], M. Harrigan[2], T. Huang[2], S. Isakov[2], E. Jeffrey[2], Z. Jiang[2], D. Kafri[2], K. Kechedzhi[2], J. Kelly[2], P. Klimov[2], A. Korotkov[2], F. Kostritsa[2], D. Landhuis[2], E. Lucero[2], A. Megrant[2], M. Mohseni[2], M. McEwen[1], J. McClean[2], X. Mi[2], J. Mutus[2], O. Naaman[2], M. Neeley[2], M. Niu[2], A. Petukhov[2], C. Quintana[2], N. Rubin[2], D. Sank[2], K. Satzinger[2], A. Vainsencher[2], T. White[2], Z. Yao[2], P. Yeh[2], A. Zalcman[2], V. Smelyanskiy[2], H. Neven[2], S. Gopalakrishnan[6], D. Abanin[7], M. Knap[3,4], J. Martinis[1,2], and P. Roushan[2]

[1]*Department of Physics, University of California, Santa Barbara, CA 93106, USA*
[2]*Google Inc., Santa Barbara, CA 93117, USA*
[3]*Department of Physics and Institute for Advanced Study, Technical University of Munich, 85748 Garching, Germany*
[4]*Munich Center for Quantum Science and Technology (MCQST), Schellingstr. 4, D-80799 München, Germany*
[5]*Department of Quantum Matter Physics, University of Geneva, Switzerland*
[6]*Department of Physics and Astronomy, College of Staten Island, Staten Island, NY 10314, USA*
[7]*Department of Theoretical Physics, University of Geneva, 24 quai Ernest-Ansermet, Switzerland*

# Contents

*These authors contributed equally



# 1 Device and calibration, Figs. S1-S3

The nearest-neighbor coupled, linear chain device used in Figs. 2 and 3 of the main text features 9 frequency tunable transmon qubits with tunable inter-qubit coupling. An optical micrograph is of this device is shown in Fig. S1 (a). The design details are discussed further in [1]. The effective circuit model for a three qubit, two coupler subsection of the device is shown in panel (b). Following [1], we infer the values of the circuit model parameters for this device from spectroscopic measurements. The dynamics of this device are described by a Bose-Hubbard Hamiltonian with tunable coefficients. We use the parameterized circuit model to create a mapping between the experimentally controlled bias currents and the resultant Hamiltonian coefficients. The circuit model measurements are made as a series of single and two qubit measurements. Once the circuit model has been developed, we benchmark the 9-qubit collective dynamics as described in Fig. S3.

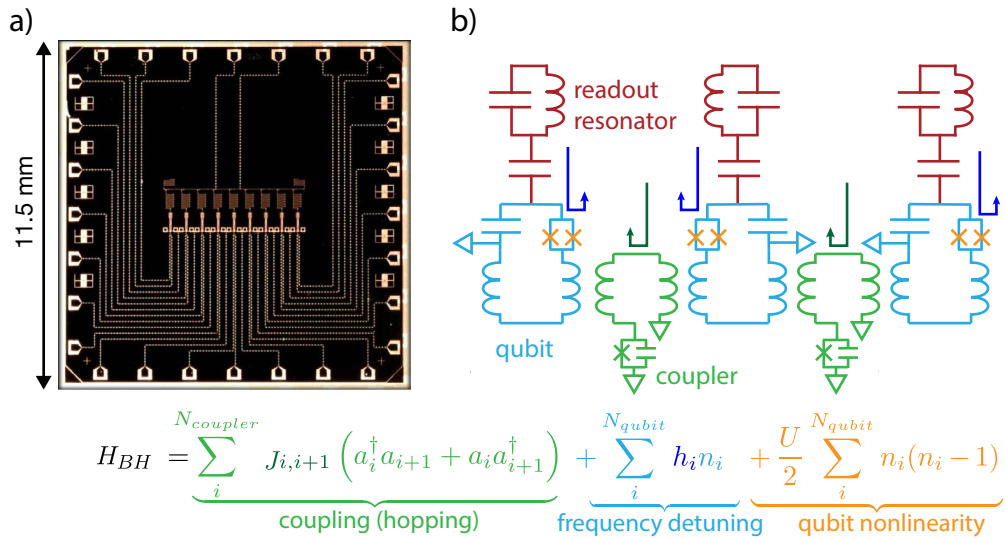

$$H_{BH} = \underbrace{\sum_i^{N_{coupler}} J_{i,i+1} \left(a_i^\dagger a_{i+1} + a_i a_{i+1}^\dagger\right)}_{\text{coupling (hopping)}} + \underbrace{\sum_i^{N_{qubit}} h_i n_i}_{\text{frequency detuning}} + \underbrace{\frac{U}{2} \sum_i^{N_{qubit}} n_i(n_i-1)}_{\text{qubit nonlinearity}}$$

**Figure S1: The 9 qubit linear chain device.** This device was used in Figs. 2-4 of the main text. **(a)** Optical micrograph of the 9 qubit linear-chain device. **(b)** Circuit diagram for a three qubit subsection of the device.

We use Clifford based randomized benchmarking (RB) and purity benchmarking to quantify the total error rate and the error rate due to decoherence per Clifford for the single qubit

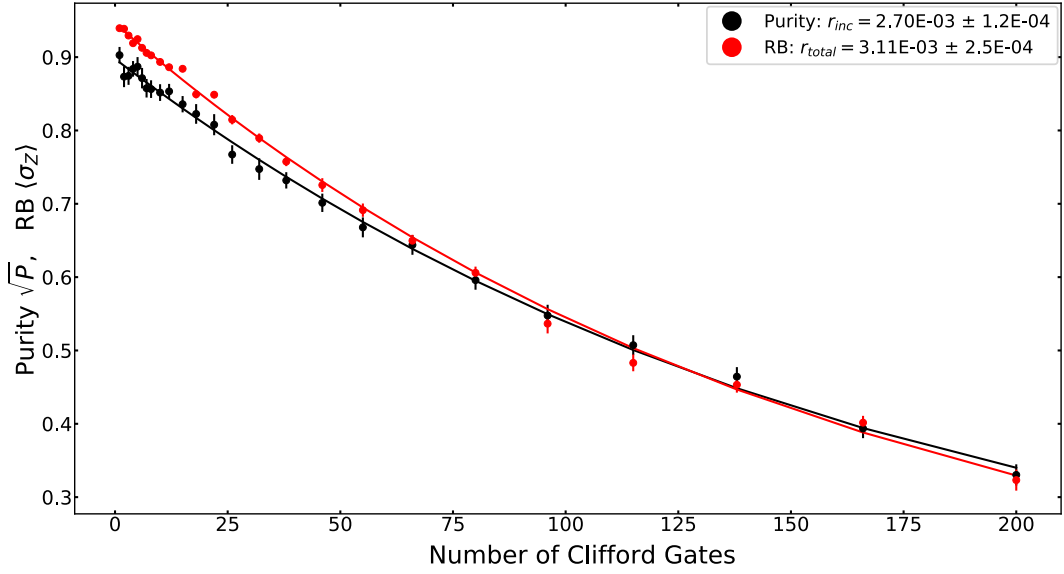

**Figure S2: Single qubit gate performance.** Clifford based randomized benchmarking (RB), shown in red, characterizes the total error rate per Clifford. Purity benchmarking, shown in black, characterizes the total incoherent error rate per Clifford.

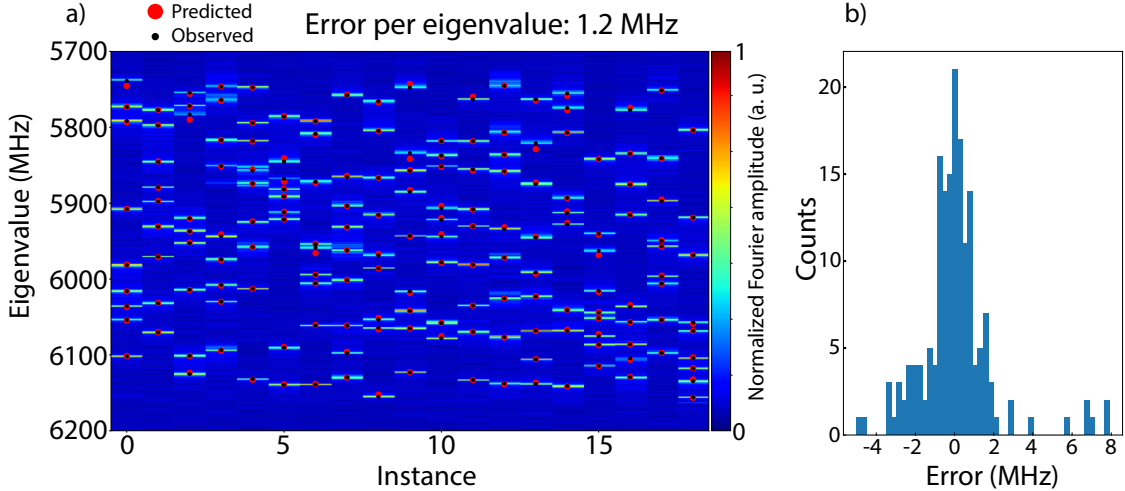

**Figure S3: Many-body Ramsey calibration data for 9 qubit randomly generated Hamiltonians.** **(a)** Fourier data for 18 instances of randomly generated Hamiltonians overlayed with the control model predictions of the eigenvalues. **(b)** A histogram of the difference between the control model predicted eigenvalues and the experimentally observed peaks.

gates in our algorithm. The data shown in Fig. S2 is for a typical qubit. The total and incoherent error rates per Clifford are extracted to be $3.1 \times 10^{-3}$ and $2.7 \times 10^{-3}$. Since there are relatively few single qubit gates in our analog algorithms this is not a significant source of error.

4

In order to benchmark our ability to set multi-qubit time-independent Hamiltonians we compare the eigenvalues predicted by our control model with those observed by using the manybody Ramsey spectroscopy technique[2]. We prepare a qubit in the superposition state $|\psi_0\rangle = \left(\frac{|0\rangle+|1\rangle}{\sqrt{2}}\right) \otimes |0, ..., 0\rangle_{\text{Other}}$, evolve the system under a 9 qubit time-independent Hamiltonian, and observe $\langle \sigma^x + i\sigma^y \rangle$ of the initialized qubit vs evolution time. The eigenvalue spectrum can then be recovered by Fourier transforming this time-series. This procedure is repeated for each of the qubits in our system and a composite spectrum is assembled from these measurements. We then compare the eigenvalues predicted by the parameterized circuit model and with those extracted experimentally. Example calibration data for the 9 qubit linear chain geometry is shown is Fig. S3.

To make a stressful test and benchmark our control model over a wide parameter space we perform manybody Ramsey spectroscopy over several instances of randomly generated Hamiltonians. In the 9 qubit data shown here, the coefficients of our target Bose-Hubbard Hamiltonian were taken to be independent random variables with $J_{ij} \in [0, 45]$ MHz and $h_i \in [-200, 200]$ MHz. In Fig. S3 (a) the 2D color map shows the composite spectra for these instances. The 2D plot is overlayed with the eigenvalue predictions from the control model (red circles) and the detected peak locations (black circles). In (b) we report the distribution of errors obtained from the difference of the predicted and observed eigenvalues for each instance. The average error per eigenvalue for these Hamiltonian instances is 1.2 MHz.

## 2  Transport measurements, Figs. S4-S6

In Fig. S4 we show data from the transport measurements before disorder averaging. The data shown is for $n_{ph} = 2$ and selected values of disorder parameter $w$ for $J = 40$ MHz. The disorder averaged data (black lines) is contained in Fig. 2 (a) of the main text, and the histograms in Fig. 2 (b) of the main text are time slices of this data at 100 ns. The spread in values at short time is primarily due to readout error, as state preparation error is small.

In the main text we report on short-time dynamics $t \lesssim 100$ ns, before our system is dominated by decoherence. In reality, our 9 qubit chain is an open system, subject to both

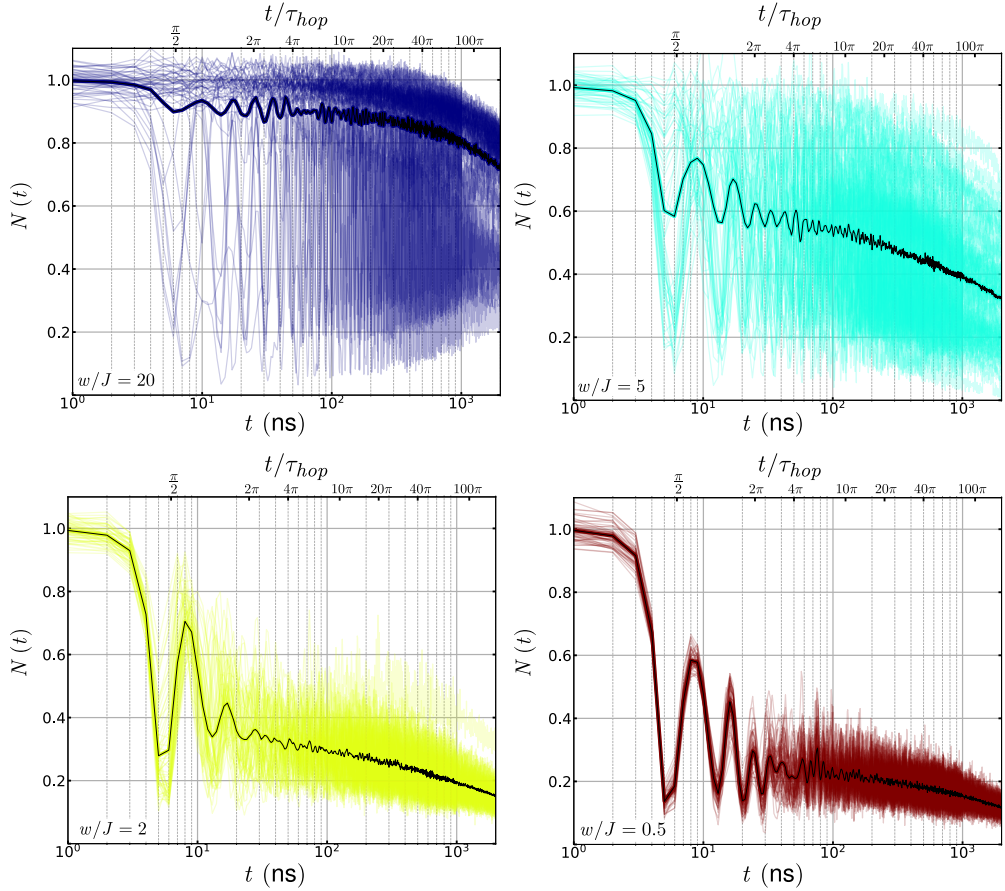

**Figure S4: Transport measurement instances.** Instances of $N(t)$ for the transport protocol of main text Fig. 2 prior to disorder averaging. Data shown here is for $J = 40\,\text{MHz}$ and $n_{ph} = 2$.

relaxation and dephasing because of its coupling to the environment. The characteristic relaxation time $T_1$ is $\sim 10\mu s$ and the characteristic dephasing time is a few $\mu s$. In Fig. S4 we provide additional data as an estimate of the importance of these open system effects. In panel (a) we show the disorder averaged population vs time data for $n_{ph} = 1$. In panel (b) we show the population vs time data for $n_{ph} = 1$ after correcting for relaxation (photon loss) using a simple single qubit $T_1$ model $\overline{N^{corrected}(t)} = \overline{N(t)}/e^{(-t/10\mu s)}$. At high disorder, where the localization length is shorter than one lattice site, single qubit $T_1$ (photon loss to the environment) is the dominant mechanism by which a photon leaves the observation site and this correction works well, as indicated by the fact that the population has taken a stationary value. At low disorder, in the diffusive regime, the excitations are able to distribute themselves evenly across the chain and we expect the $T_1$ correction to work well in

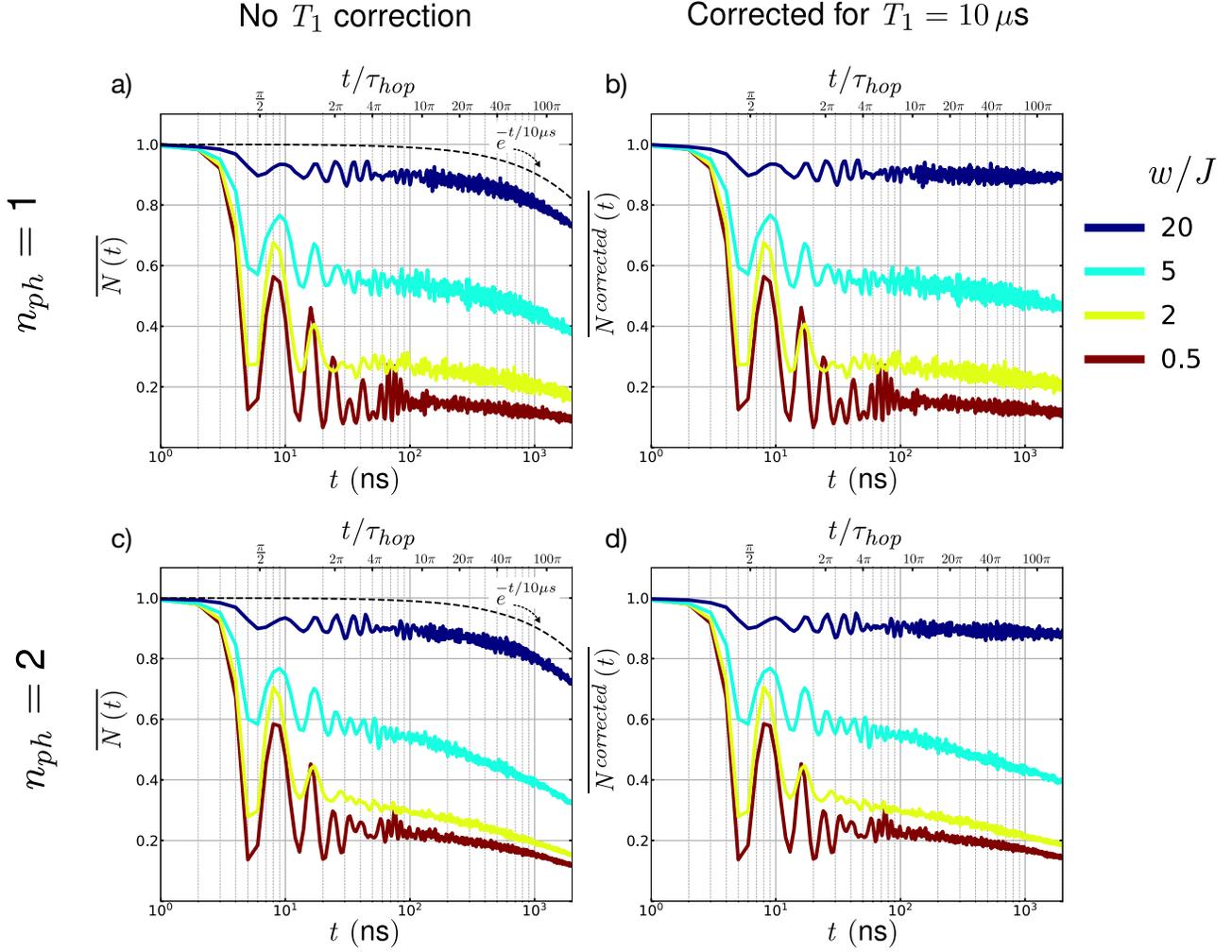

**Figure S5: Transport Measurements: Decoherence Effects.** **(a)** The raw disorder averaged transport data for $n_{ph} = 1$ **(b)** The data from (a) corrected for a simple energy relaxation model $\overline{N^{corrected}(t)} = \overline{N(t)}/e^{(-t/10\mu s)}$. **(c)** The raw disorder averaged transport data for $n_{ph} = 2$. **(d)** The disorder averaged transport data with $T_1$ correction.

this case as well. Referring to Fig. 2 (c) of the main text we see that at 100 ns in the diffusive regime at low disorder we measure the thermal expectation values. This indicates clearly that relaxation effects are not significant in the first 100 ns. And that any apparent loss is due to transport within the 9 qubit chain and not photon loss. For intermediate disorders there appears to be additional photon loss since the onsite population declines. However, the decrease in observed population at the observation site is attributed to dephasing assisted delocalization.[3–7]

When the LIOM extends over multiple lattice sites, dephasing between the sites breaks down the localized wave-packet by destroying the quantum interference pattern that causes the localization. This breakdown of coherence between different parts of the wave packet enables transport of the excitation across the 9 qubit chain. Crucially, we note that neither $T_1$ relaxation nor dephasing between the lattice sites significantly influence the dynamics at higher disorders or short times. This feature is captured in the main text Fig. 4 (b) where we note that $\overline{\langle \sigma^z \rangle}$ is nearly constant between 10 ns and 100 ns. In Fig. S5 (c) and (d) we show the raw and $T_1$ corrected data for $n_{ph} = 2$.

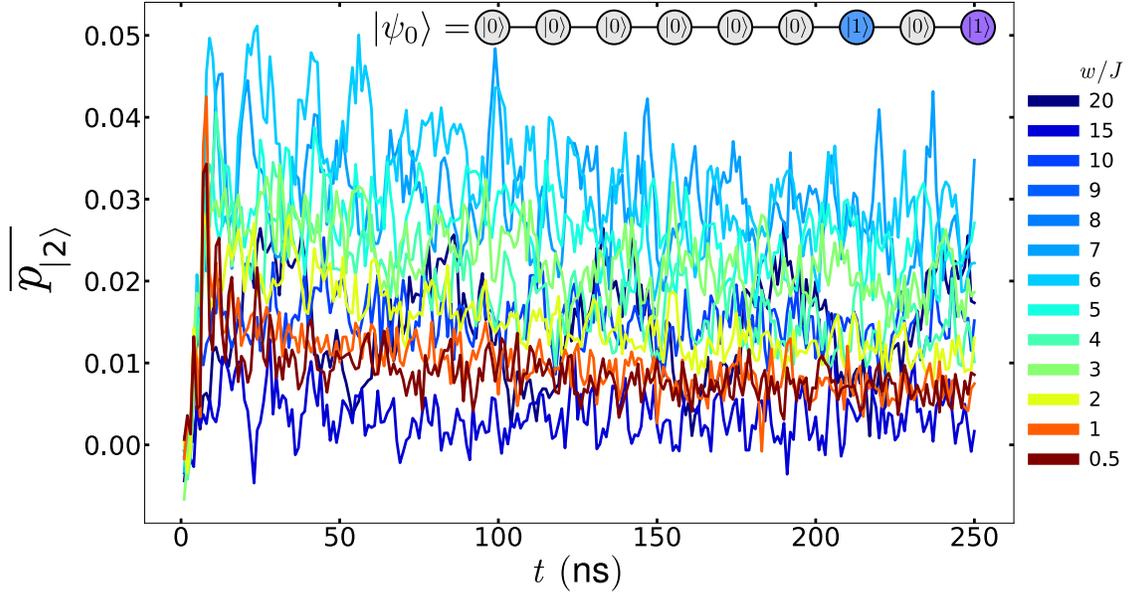

Figure S6: **Transport Measurements: Two state occupation.**

A critical feature of our system is that multiple excitations in the system may interact via the Hubbard interaction. The form of this interaction $H_{int} = \frac{U}{2} \sum_{n=1}^{n_Q} a_n^\dagger a_n (a_n^\dagger a_n - 1)$ indicates that it is only activated when there are multiple excitations on the same lattice site. Thus the interaction effects that we report in the main text require occupation of the higher levels of our Bose-Hubbard lattice. In Fig. S6 we report the $|2\rangle$ population vs time for a system initially in the state $|\psi_0\rangle = |000000101\rangle$ and observed on the right-most qubit. We find that

8

the $|2\rangle$ state population is typically at the 2 % level, achieves its maximum value early in the evolution, and does not progressively grow larger with time.

## 3 Interferometric protocols, Figs. S7-S10

In order to gain some insights about the echo sequences, we first consider the case of very strong disorder, where the local integrals of motion (LIOMs) $\tau_i^z$ are close to the physical spins $S_i^z$ (represented by the two lowest energy levels of a qubit), and assume that we directly manipulate LIOMs. First, we will consider the spin echo sequence illustrated graphically in Fig. S7 [8]. Assuming we start from the vacuum state $|\psi_0\rangle = |0\rangle \otimes |\{\tau_j\}\rangle$, we initiate the dynamics by applying a $\pi/2$ pulse:

$$|\psi\rangle = \frac{1}{\sqrt{2}}(|0\rangle + i|1\rangle) \otimes |\{\tau_j\}\rangle \tag{1}$$

When the system evolves for times $t/2$, the spin at site $i$ experiences an effective magnetic field, that depends on the states of the other LIOMs, see Eq. (2) in the main text,

$$\Delta_i = \tilde{h}_i + \sum_j J_{ij}\tau_j^z + \sum_{j,k} J_{ijk}\tau_j^z\tau_k^z + \ldots. \tag{2}$$

The $\pi$ rotation halfway through the spin-echo sequence then inverts the effective magnetic

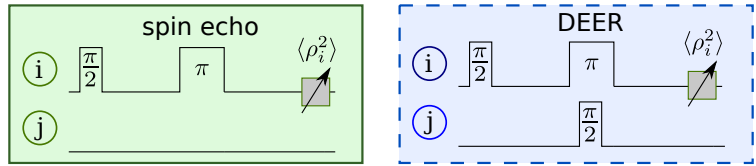

**Figure S7: Pulse sequence schematics for spin and DEER echo.** DEER echo differs from spin echo by the addition of a remote $\pi/2$ pulse simultaneous with the spin echo $\pi$ pulse between the free precession intervals.

field $\Delta_i \to -\Delta_i$ which is precisely canceled after another time evolution for $t/2$. At the end of the protocol we measure the purity, which is advantageous over measuring a single spin component, because it is less prone to running field gradients and external perturbations. For

the spin-echo sequence on the LIOMs we find a perfect purity of one. In a true measurement on our device the echo is performed on the physical spins, which possess a finite operator overlap with the LIOMs which is less than one. This leads to a spin echo signal that saturates to a finite value that decreases with decreasing disorder strength [8].

In the DEER echo sequence we similarly perform a spin echo measurement on site $i$ as before, However, half-way through the time evolution we modify a second part of the system, say site $j$ by applying a $\pi/2$ pulse, see Fig. S7. The effective magnetic field for the backward evolution $\tilde{\Delta}_i$, deviates from the field $\Delta_i$ of the forward evolution in all the terms containing $\tau_j^z$. In summary, the state after the second time evolution is therefore

$$|\psi_D(t)\rangle = \frac{1}{2}[\,|1\rangle \otimes |...0_j...\rangle + e^{i(\Delta_i - \tilde{\Delta}_i)t - i\Delta_j t} \,|1\rangle \otimes |...1_j...\rangle - \\ |0\rangle \otimes |...0_j...\rangle - e^{-i(\Delta_i - \tilde{\Delta}_i)t - i\Delta_j t} \,|0\rangle \otimes |...1_j...\rangle] \tag{3}$$

and the measurement of the purity then yields

$$\text{tr}\left(\rho^2\right) = \cos^2\left[\left(\Delta_i - \tilde{\Delta}_i\right)t\right]. \tag{4}$$

Due to the interaction between the $\tau$ bits at site i and j, the phases do not cancel anymore and the signal decays. The difference between spin and DEER echo is thus a pure interaction effect which would not appear in the noninteracting localized phase. The advantage of performing a differential measurement of the two echo protocols is that even in the presence of noise, deviations of the two echo signals, demonstrates a clear interaction effect and hence is able to unambiguously measure the interacting character of the LIOMs. Because these interaction effects are due to the local occupation of higher orbitals we numerically estimate the population of multiply excited states $n_i^{max} = 1, 2, 3$ during the DEER echo protocol for a evolution time of $t = 63\,\text{ns}$ in Fig. S8.

In the experimental measurement of the purity, local occupations higher than two are not taken into account. This leads to a leakage of the measurement as characterized by the finite value of $\langle \sigma_i^z \rangle$ in Fig. 3 (b) of the main text. Moreover, we numerically estimate this

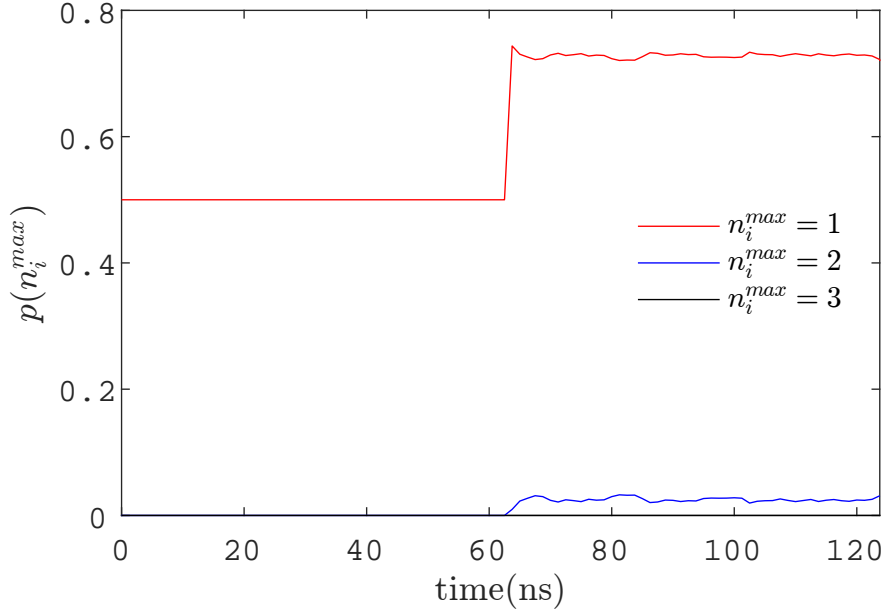

**Figure S8: Numerical estimate of the occupation of higher transmon levels during the DEER echo protocol.** Calculation shown here for for the one, two, and three excitation states

effect by resolving the probabilities for maximum local occupations $n_i^{max} = 1, 2, 3$ during the DEER echo protocol for a evolution time of $t = 63\,\text{ns}$ in Fig. S8. From that it can be deduced that leakage effects are not severe, and in particular does not change the qualitative difference between the spin echo and DEER echo protocols. Probability for maximum local occupations of $n_i^{max} = 1, 2, 3$ during the DEER echo protocol for $L = 9$ sites, an evolution time of $T = 63\text{ns}$, coupling $J = 2\pi \cdot 40\text{MHz}$, disorder strength $w/J = 10$ and interaction $U/J = 4$.

In Fig. S9 we compare the data from the interferometric pulse sequences presented in Fig. 3 of the main text with numeric predictions. In panels (a) and (b) we compare the onsite population at the spin echo qubit. Although there is a strong correspondence, we observe greater diffusion off site (larger $\sigma^z$)) in the experiment than in the numerics. There is also less contrast in the experiment than in the numerics. It is likely that these differences are related to the transient pulse response of our system and open systems effects, however further investigation is needed to make a conclusive determination.

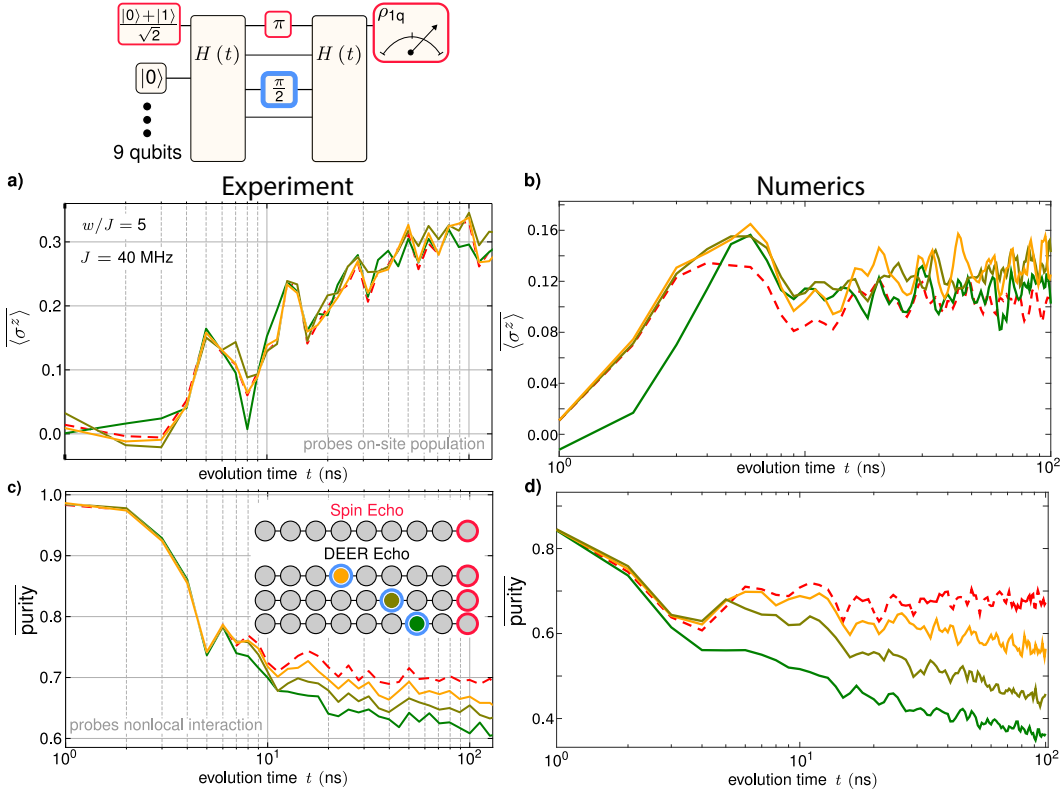

**Figure S9: Comparison with numerics for echo experiments**. **(a)** and **(b)** Disorder averaged expectation value of $\sigma^z$ showing population at the observation site for experimental observations and numerical prediction. The lack of dependence of $\overline{\langle \sigma^z \rangle}$ on the DEER pulse indicates localization in our system. **(c)** and **(d)** Disorder averaged purity of the reduced single qubit density matrix at the observation site. The contrast between spin-echo and DEER echo demonstrates that the local phase accumulation is conditional on the remote population. This is a direct measure of the nonlocal interaction strength.

In Fig. S10 we show extended data for echo sequence measurements for several values of of the disorder parameter $w$ with $J$ held fixed at 40 MHz. Compared with Fig. 3 of the main text, the initial state for these measurements had an additional excitation at the indicated position (purple). We observe a strong interferometric signature in the purity, indicating nonlocal interaction. In these measurements $\sigma^z$ does not depend on the position of the echo pulses, indicating localization.

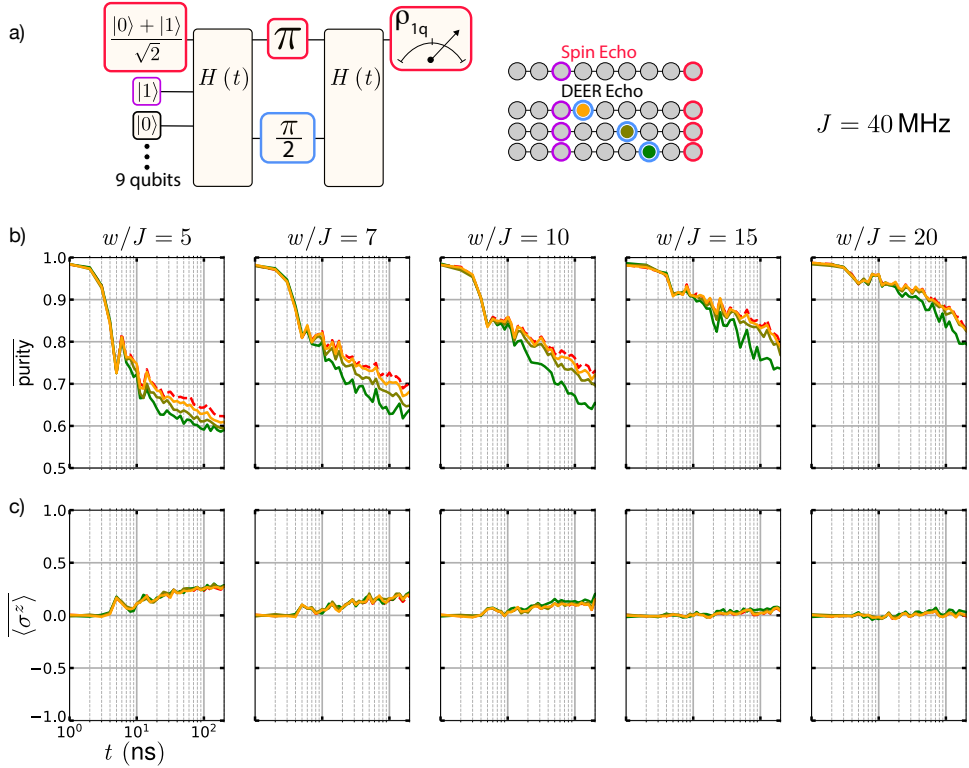

**Figure S10: Interferometric Protocols: Extended Data.** **(a)** Spin and DEER echo pulse sequences. We the blue outline indicates the position of the DEER echo pulse, and the position of an additional excitation is indicated in purple. **(b)** purity of the single qubit density matrix after the spin echo (dashed red lines) and DEER echo (solid lines) experiments. **(c)** $\langle \sigma^z \rangle$ monitored over the echo experiments.

# 4 Entanglement measures

The distillable entanglement of the two qubit density matrix $E_D(\rho_{2q})$ is lower bounded by the coherent information entropy

$$E_D(\rho_{2q}) \geq S(\rho_{1q}) - S(\rho_{2q}), \tag{5}$$

where $\rho_{1q,2q}$ are the reduced density matrices of one of the two qubits and the two qubit subsystem, respectively, and $S(\rho)$ is the von Neumann entanglement entropy. An upper bound to the distillable entanglement is provided by the logarithmic negativity[9] which is defined as

$$E_N(\rho_2) = \log_2 ||\rho_2^{T_A}||_1. \tag{6}$$

13

Here, $\rho_2^{T_A}$ is the partial transpose of the reduced density matrix with respect to one of the qubits and $||\cdot||_1$ denotes the trace norm.

A second operational entanglement measure is the entanglement of formation, which is a proxy for entanglement cost, i.e. the amount of entanglement needed to create a given entangled state. It is defined as

$$E_F(\rho) = \epsilon(\mathcal{C}(\rho)) \tag{7}$$

with

$$\epsilon(x) = -h_+(x)\log_2 h_+(x) - h_-(x)\log_2 h_-(x) \tag{8}$$

where

$$h_\pm(x) = -\frac{1}{2}\left(1 \pm \sqrt{1-x^2}\right). \tag{9}$$

The concurrence $\mathcal{C}(\rho)$ of a mixed state of two qubits is defined as

$$\mathcal{C}(\rho) = \max\left(0, \lambda_1 - \lambda_2 - \lambda_3 - \lambda_4\right), \tag{10}$$

where $\lambda_i$ are the eigenvalues of

$$R = \sqrt{\sqrt{\rho}\tilde{\rho}\sqrt{\rho}} \tag{11}$$

and

$$\tilde{\rho} = (\sigma_y \otimes \sigma_y)\rho^*(\sigma_y \otimes \sigma_y). \tag{12}$$

## 5  Extended data for 1D qubit array, Fig. S11 - S12

A hallmark of the MBL phase is the slow growth of entanglement, contrasting with Anderson localization where the entanglement is constant. To study the development of entanglement entropy, we designate two qubits as a subsystem and the rest of the chain as the environment (Fig. S11(a)), and directly measure the evolution of the reduced density matrix of the subsystem. The subsystem qubits are initialized into superposition states. Fig. S11(b) shows that $\overline{\langle \sigma^z \rangle}$ initially rises because population from the subsystem qubits is transferred to the

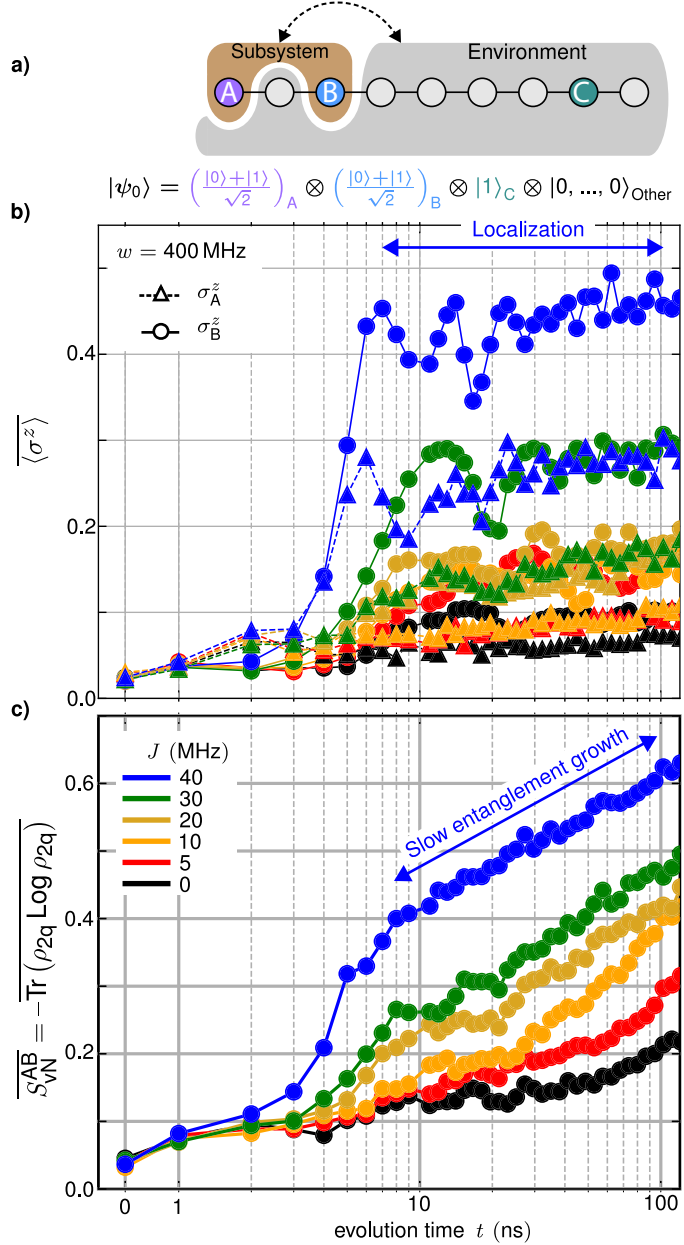

**Figure S11: Localization and slow growth of entanglement (a)** Partitioning of our 9-qubit chain into a subsystem and environment. The subsystem qubits (A and B) are initialized into superposition states, and the system is loaded with an additional excitation (site C) to enhance many-body interactions. We tomographically reconstruct the density matrix of the subsystem. **(b)** $\overline{\langle \sigma^z \rangle}$ for subsystem qubits. **(c)** von Neumann entanglement entropy of the two qubit subsystem for several coupling strengths.

environment which has a smaller photon density. After this initial rise, $\overline{\langle \sigma^z \rangle}$ takes a stationary value which decreases with decreasing coupling strength, establishing the localization of our system.

15

We use the von Neumann entanglement entropy

$$S_{\text{vN}} = -\text{Tr}\rho_{2\text{q}}\log\rho_{2\text{q}} \qquad (13)$$

to quantify the entanglement between the subsystem and the environment (panel (c)). The initial increase in $\overline{S_{\text{vN}}}$ occurring simultaneously with the increase in $\overline{\langle\sigma^z\rangle}$ is understood as the result of the subsystem exchanging population with the environment. Thereafter, while the system is demonstrably localized, we observe logarithmic growth of von Neumann entropy. We can understand the slow growth in terms of the LIOM framework: The non-local and exponentially decaying interactions between the LIOMs give rise to dephasing between the qubits and follow a broad log-normal distribution [10]. As a consequence, the entanglement of individual runs is strongly fluctuating on different time scales leading to a logarithmic growth of the entanglement of the subsystem. We note that preparing subsystem qubits in an x-polarized state is key for the success of this measurement as it enhances the measurement visibility by being highly phase sensitive. The von Neumann entropy quantifies entanglement with all external degrees of freedom and is not able to disambiguate entanglement with the environmental qubits due to unitary dynamics from open systems effects. As such, our observed entropy is an upper bound on the entanglement generated within our qubit array. The $J = 0$ curve (black) provides an estimate of the amount of entropy that is due to open system effects. Next, we introduce entanglement measures that are more robust against open systems effects and lower bound the entanglement between parts of the system.

In a 1D system we investigate the formation and preservation of entanglement between two qubits A and B that are embedded in a many-body localized environment as illustrated in Fig. S12 (a) and contrast this behavior with a system in the diffusive regime. The entanglement of formation quantifies the amount of entanglement directly between qubits A and B that would be required to produce the observed two-qubit mixed state density matrix.In panel (b), we initialize the sub-system into an unentangled product state of single qubit superpositions and observe the development of entanglement between our sub-system qubits. At high disorder, associated with the localized phase, entanglement grows continuously be-

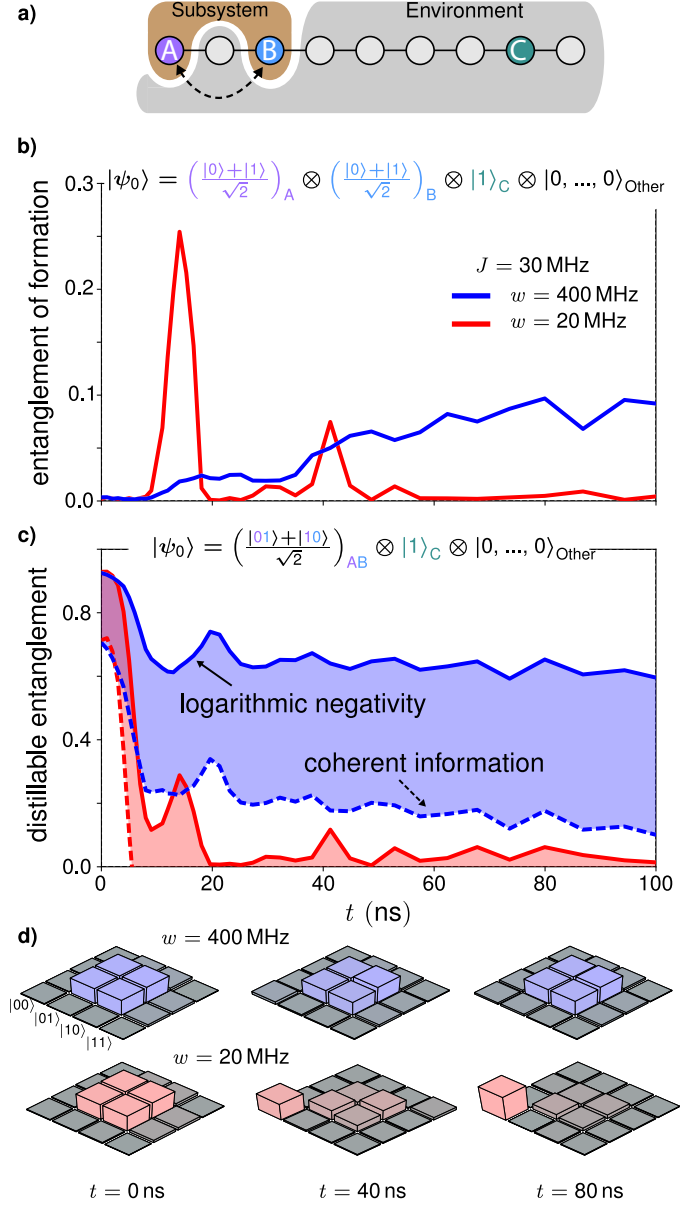

**Figure S12: Entanglement of formation and distillable entanglement in MBL and diffusive regimes** (a) Schematic diagram emphasizing our focus on the entanglement between qubits A and B which are embedded in an environment. (b) To observe the development of entanglement between sites A and B the sub-system is initialized in a product of single qubit superposition states and the entanglement of formation of the two qubit density matrix is extracted. (c) We demonstrate the capability of the MBL phase to preserve entanglement initializing the sub-system into a maximally entangled Bell pair and observing the decay of quantum correlations. We extract the logarithmic negativity and coherent information from measurements of the two-qubit sub-system density matrix. These provide, respectively, upper and lower bounds on the distillable entanglement within the sub-system. (d) Representative density matrices from single disorder instances contained in (c) at high and low disorder.

tween the spatially separated sites. At low disorder, corresponding with the ergodic phase, we observe brief intervals of significant entanglement as the excitations delocalize across the

17

full 9 qubit system. However, this behavior is quickly damped as the excitations are absorbed by environmental qubits, as the full 9-qubit system thermalizes.

Systems in the MBL and diffusive regimes also differ in their ability to retain correlation between their constituent parts. This is illustrated in Fig. S12 (B) where we prepare a distant Bell state between the first and the third qubit and study the entanglement dynamics. While dephasing between LIOMs will ultimately destroy the entanglement, it will only due so on exponentially long times due to the localization. Crucially, the subsystem is in a mixed state, because it is coupling to the other 7 qubits of our device. We therefore characterize the entanglement of the 2-body mixed density matrix $\rho_{2q}(t)$ using an operational entanglement measure. In particular, we focus on the distillable entanglement, i.e., the entanglement which can be extracted from the mixed density matrix, that is upper bounded by logarithmic negativity entropy and lower bounded by the coherent information entropy. These bounds are shown in panel (c). For weak disorder (red), the prepared quantum information is immediately lost because the quantum dynamics entangles the subsystem with its environment and a featureless high temperature state is attained locally. This behavior can also be understood in terms of the monogamy of entanglement.[11] Although the two qubit subsystem is initially prepared in a maximally entangled state the degree of quantum correlation between subsystem sites decreases as the subsystem exchanges information with the environment and entangles with it. This monogamic principle also explains the damping of the peak in the low disorder data of panel (b). However, for strong disorder (blue) the distillable entanglement is sizable over long times, and hence the density matrix can be used as quantum resource. This is exemplified in panel (d) which shows the tomographic reconstruction of the two qubit density matrix for a single disorder instance as it evolves in time. These results show that a many-body localized system can efficiently retain quantum information over long time scales.

## 6 Extended data for 2D qubit arrays, Fig. S13

In Fig. S13, we show extended data for the onsite population for 2D geometries, for $n_{ph} = 1, 2, 3, 4$. The initial location of the excitations was randomized between runs but

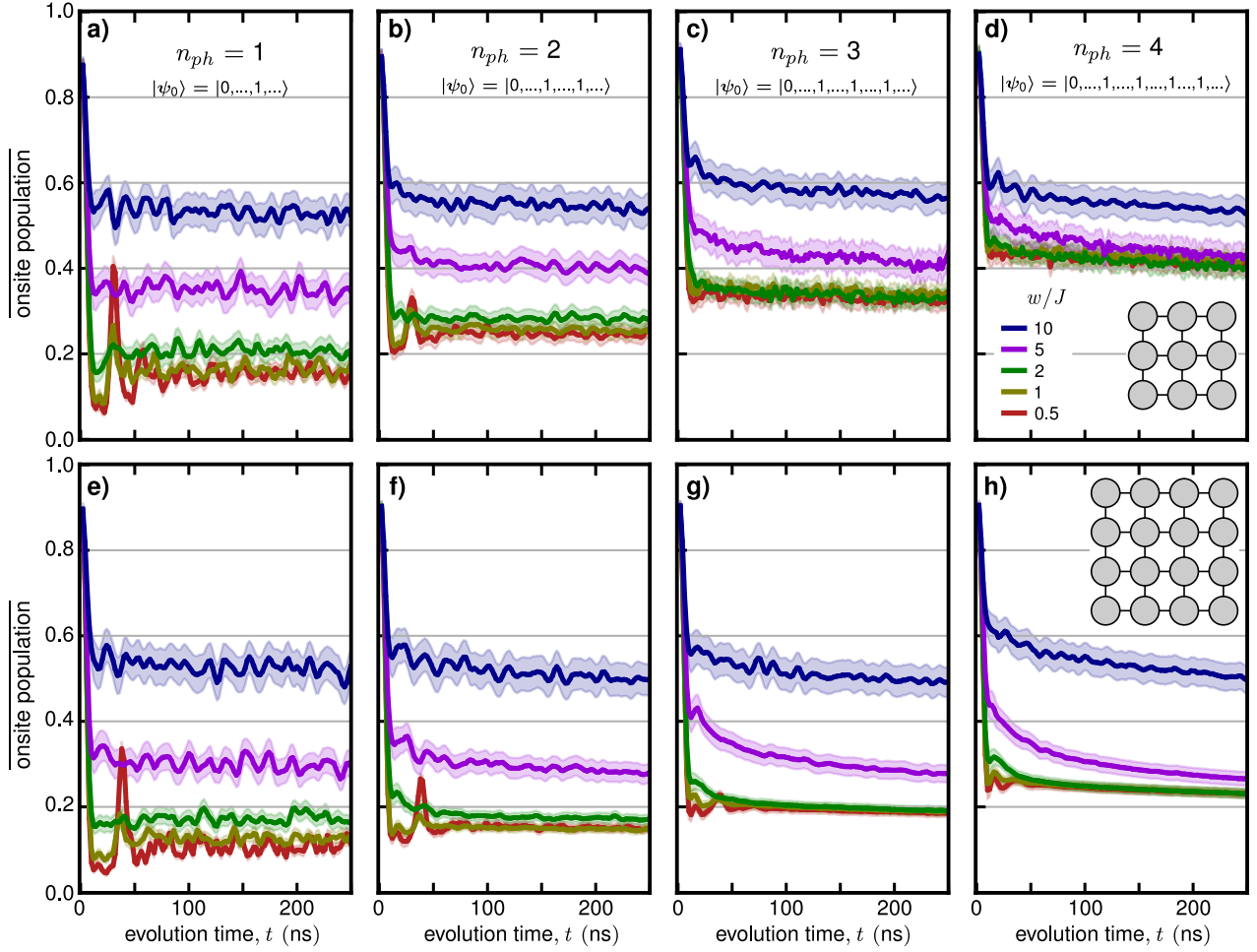

**Figure S13: Extended data for onsite population of 2D arrays.** **(a-d)** Onsite population for $n_{ph} = 1, 2, 3, 4$ on a 3x3 array of qubits. **(e-h)** Onsite population for $n_{ph} = 1, 2, 3, 4$ on a 4x4 array of qubits.

the observation site was always one of the initially excited qubits. Similar to the 1D geometries, with sufficient disorder the onsite population takes a non-thermal stationary value and is consistent with many-body localization. In the 2D geometries the onsite population consistent with thermalization at higher disorders when there are more photons in the system (greater $n_{ph}$), as in the 1D case.

## 7 Density matrix evolution numerics comparison, Figs. S14 - S18

In Fig. S11 we observe the entropy accumulation of an x-polarized subsystem in an MBL environment. The von Neumann entropy represents contributions from entanglement within

19

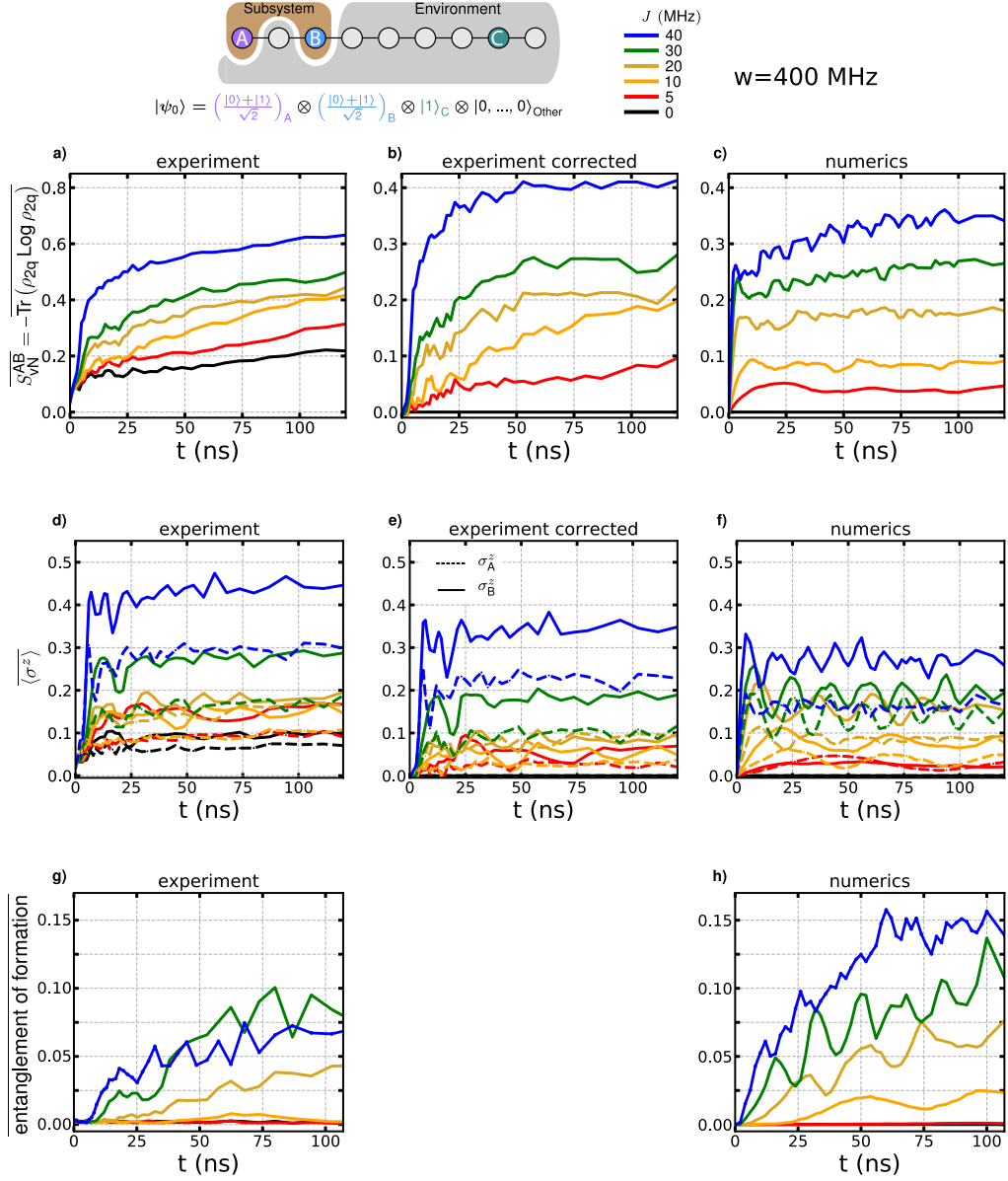

**Figure S14: Entropy Comparison with Numerics** **(a)** Raw experimental observations from the two qubit density matrix measurements. The $J=0$ data acts as a control experiment. We attribute entropy accumulation in the control experiment to open system effects **(b)** Experimental data after subtracting the baseline entropy measured in the control experiment from each of the data series. **(c)** Result of exact diagonalization numerics. **(d)** Raw experimental observation of $\sigma^z$, quantifying population. For the $J=0$ baseline case we attribute the non-zero $\sigma^z$ to state initialization error, and relaxation processes T1. **(e)** Experimental data corrected by subtracting the value of $\sigma^z$ in the $J=0$ control experiment from each of the data series. **(c)** Prediciton from exact diagonalization numerics. **(g)** Entanglement of formation as observed in the experiment. This is an affirmative observation of quantum correlation between sites A and B which cannot be attributed to open system effects, in contrast to the von Neumann entropy. The EOF observed in the experiment is slightly damped due to open system effects.

20

the 9 qubit system, as well as from open system effects. In Figs. S14-S17 we provide supporting information to assist the reader in estimating the role of open systems effects in our experiment. We find that a good estimate of the contribution to the von Neumann entropy coming from coupling to the open system is provided by the entropy of the $J = 0$ curve. In the $J = 0$ case we do not expect interaction with the environmental qubits and attribute observed entropy in that case to extrinsic dephasing and relaxation processes.

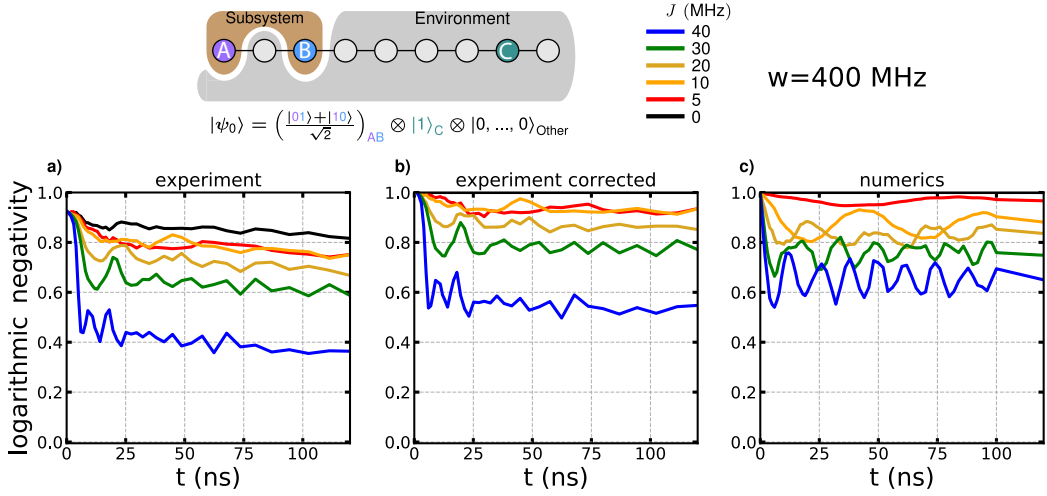

**Figure S15: entropy comparison with numerics (a)** logarithmic negativity as observed in the experiment. the black $J = 0$ curve is our control experiment, and departure of the logarithmic negativity from 1 is attributed to state initialization error and open systems effects. **(b)** experimental data corrected for the loss of correlation observed in the $J = 0$ case. The correction was performed by adding (1 - logarithmic negativity($J = 0$) ) to each data series. **(c)** prediction from exact diagonalization numerics.

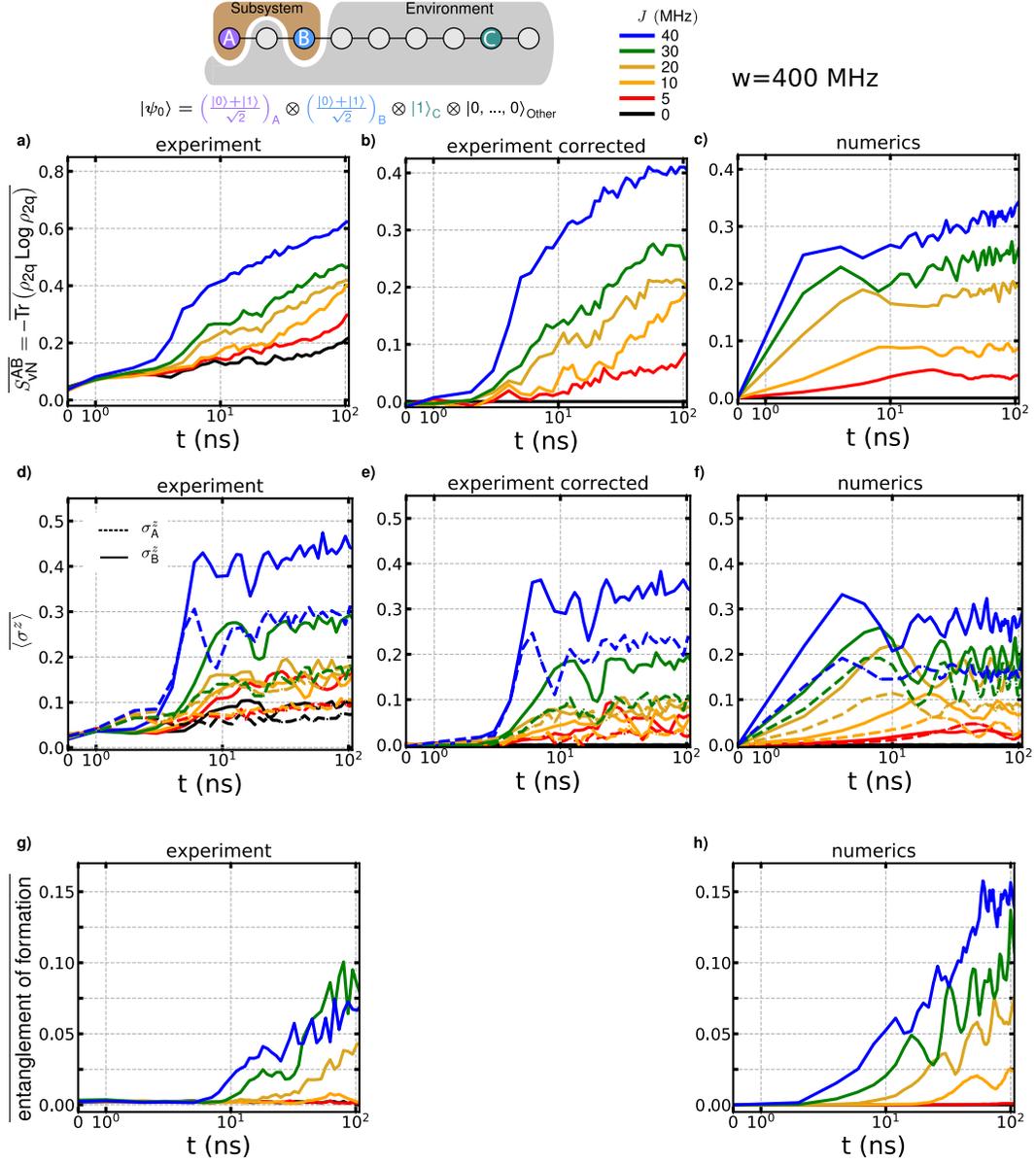

**Figure S16:** Data from Fig. S11 plotted on semi-log axes to emphasize scaling. The disagreement at short times is attributed to the transient respose of the control pulses.

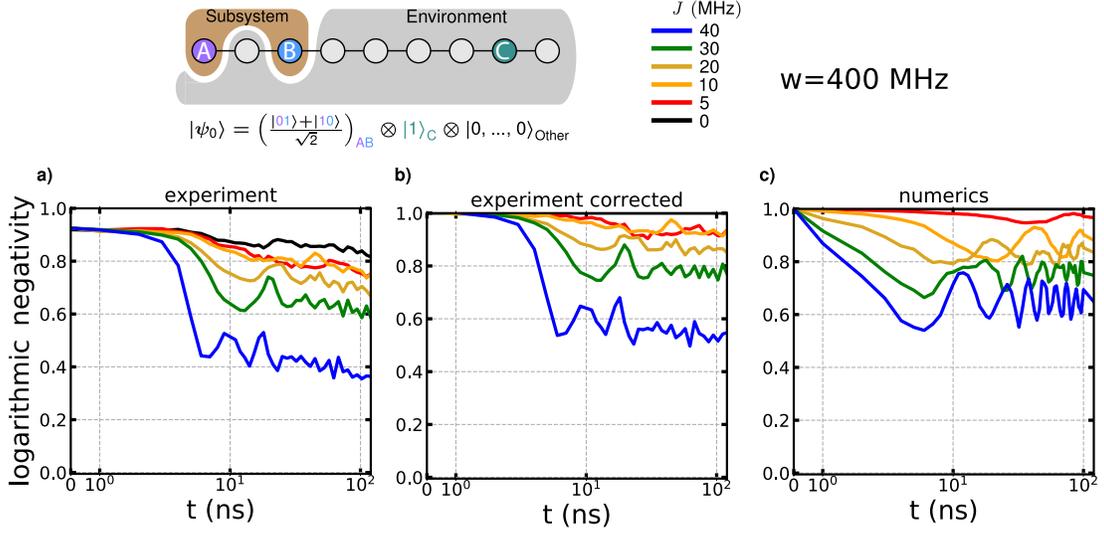

**Figure S17:** Data from Fig. S12 plotted on semi-log axes to emphasize scaling. The disagreement at short times is attributed to the transient respose of the control pulses.

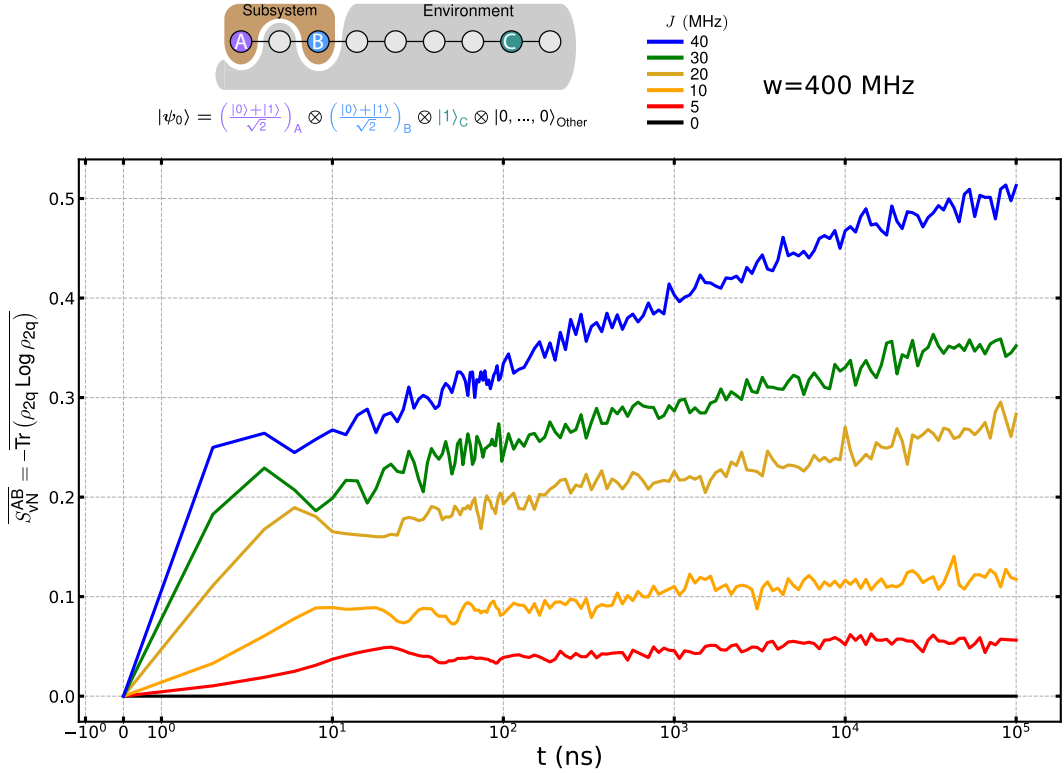

**Figure S18: Entropy comparison with numerics.** Numerics to longer times than are accessible in the experiment illustrating the predicted logarithmic growth of entanglement for our system.

23

# 8 Sensitivity to nonlinearity $U$, Figs. S19-S21

The Hamiltonian parameter $U$ varies weakly as a function of the qubit frequencies and inter-qubit coupling. $U$ cannot be controlled independently in our system. Here we provide numerical evidence that the dynamics that we report in the MBL regime are not sensitive to this parameter.

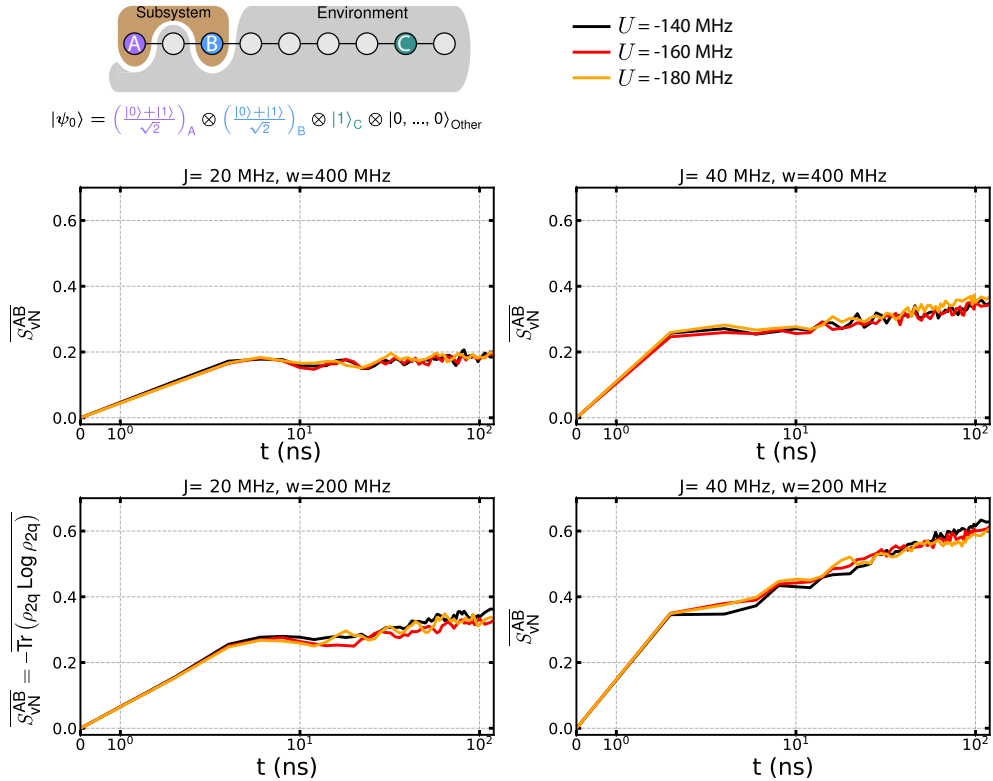

**Figure S19: Disorder averaged von Neumann entropy vs. $U$ for selected couplings and disorder magnitudes.** The von Neumann entropy observed in the experiment is predicted to be insensitive to the precise value of $U$.

24

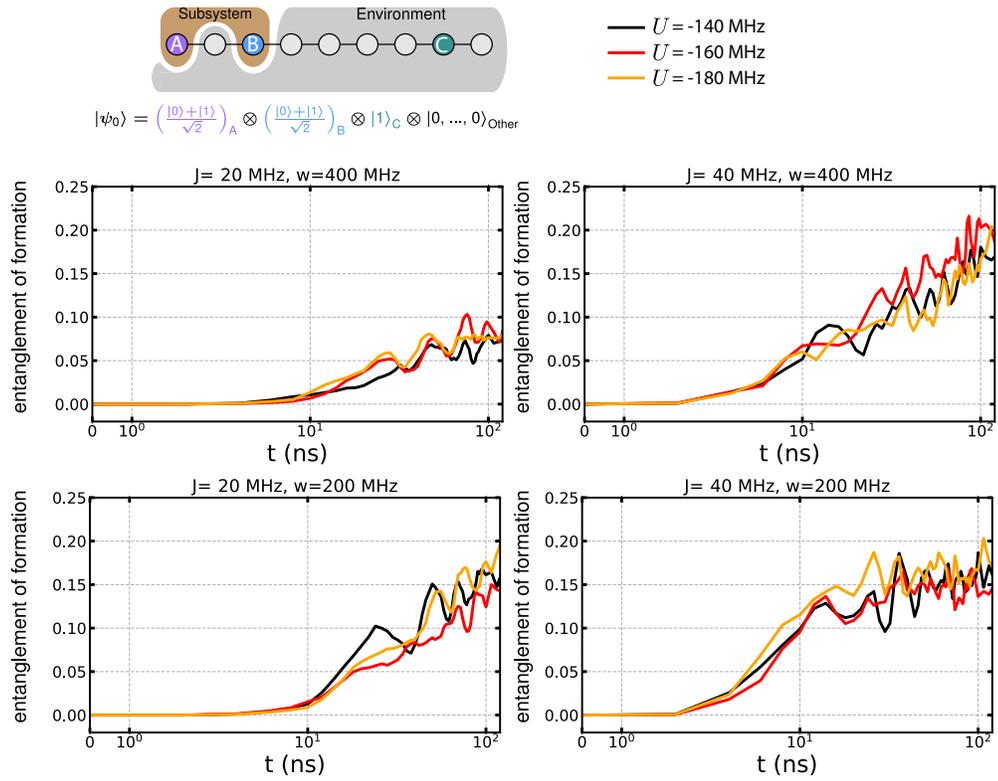

**Figure S20: Disorder averaged entanglement of formation vs. $U$ for selected couplings and disorder magnitudes.** The entanglement of formation observed in the experiment is predicted to be insensitive to the precise value of $U$.

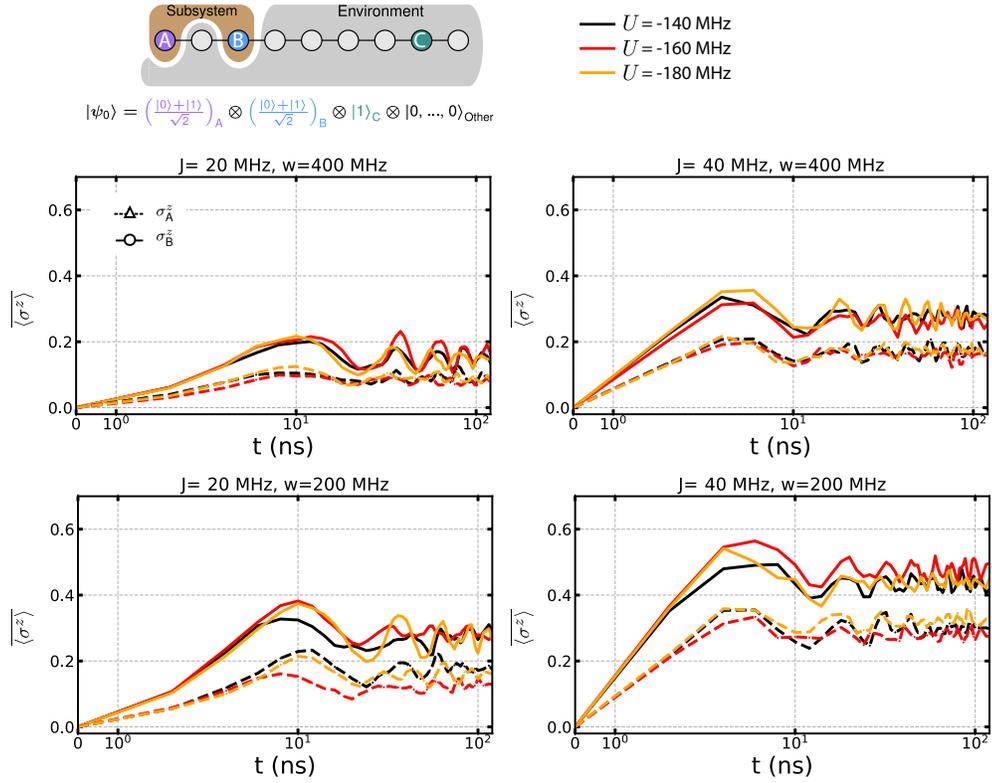

**Figure S21:** $\overline{\langle \sigma^z \rangle}$ **vs.** $U$ **for selected couplings and disorder magnitudes.** The onsite population observed in the experiment is predicted to be insensitive to the precise value of $U$.

## 8.4 Extended data - Long time behavior from superposition state

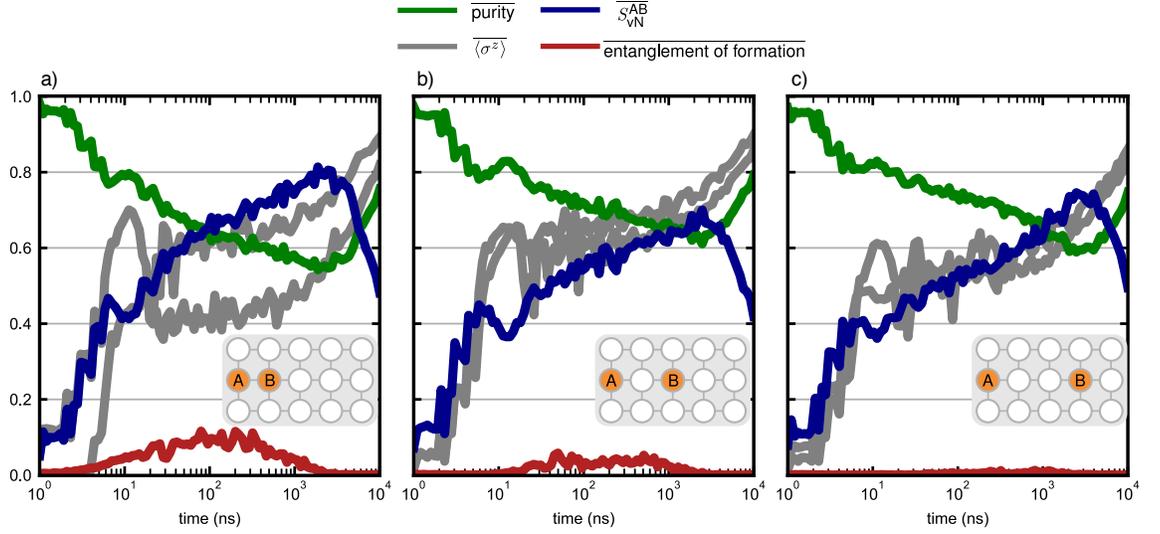

**Figure S22: Extended data for superposition state evolution in a 2d system.** At short times $\overline{\sigma^z}$ (grey) is relatively flat after its initial jump and the entanglement measures $\overline{S_{vN}^{AB}}$ (blue) and Entanglement of formation (red) reflect entanglement growth within the qubit system. The sharp increase in $\overline{\sigma^z}$ at long time indicates that the system is relaxing to the ground state. For this reason $\overline{S_{vN}^{AB}}$ and $\overline{\text{Entanglement of formation}}$ decrease while the purity (green) increases at long times.



\end{document}